\documentclass[sigconf]{acmart}
\AtBeginDocument{%
  }

\setcopyright{acmcopyright}
\copyrightyear{2023}
\acmYear{2023}
\acmDOI{XXXXXXX.XXXXXXX}

\acmConference[ACSAC '23]{Annual Computer Security Applications Conference}{December 4--8, 2023}{Austin, Texas}
\acmPrice{15.00}
\acmISBN{978-1-4503-XXXX-X/18/06}

\usepackage[ruled,linesnumbered]{algorithm2e}
\usepackage{amsfonts} 
\usepackage{amsmath}
\usepackage{mdwmath}
\usepackage{nicefrac}
\usepackage{caption}% caption=false
\usepackage[font=footnotesize]{subfig}
\usepackage{xcolor}

\usepackage{multirow}
\usepackage{url}
\usepackage{comment}
\usepackage{soul}

\usepackage{enumitem}
\DeclareMathAlphabet{\mathcal}{OMS}{cmsy}{m}{n}

\newcommand{\ournameNoSpace}{FLEDGE}
\newcommand{\ourname}{\ournameNoSpace\xspace}
\newcommand{\ournameGen}{\ournameNoSpace's\xspace}
\newcommand{\adversaryNoSpace}{\ensuremath{\mathcal{A}}}
\newcommand{\adversary}{\adversaryNoSpace\xspace}
\newcommand{\adversaryPrivacy}{\ensuremath{\adversary^p}\xspace}
\newcommand{\adversaryPoisoning}{\ensuremath{\adversary^s}\xspace}
\newcommand{\aggregationServer}{\ensuremath{\mathcal{S}}\xspace}

\newcommand{\fedavg}{\mbox{FedAVG}\xspace}
\newcommand{\etal}{\emph{et~al.}\xspace}
\newcommand{\sect}{Sect.~}
\newcommand{\nonIidNoSpace}{non-IID}
\newcommand{\nonIId}{\nonIidNoSpace\xspace}
\newcommand{\nonIid}{\nonIidNoSpace\xspace}

\newcommand{\Sota}{State-of-the-art }
\newcommand{\sota}{state-of-the-art }
\newcommand{\polydeg}{\mbox{PolyDeg}\xspace}

\begin{document}
%don't want date printed
%\date{}
\title{\ourname: Ledger-based Federated Learning Resilient to Inference and Backdoor Attacks } 

\author{Jorge Castillo}
\email{jorge.a.castillo01@utrgv.edu}
\affiliation{
    \institution{The University of Texas Rio Grande Valley$^1$}
    \country{USA}
}
\author{Phillip Rieger}
\email{phillip.rieger@trust.tu-darmstadt.de}
\affiliation{%
  \institution{Technical University of Darmstadt}
  \country{Germany}
}

\author{Hossein Fereidooni}
\email{hossein.fereidooni@kobil.com}
\affiliation{%
  \institution{KOBIL GmbH$^2$}
  \country{Germany}
}

\author{Qian Chen}
\email{guenevereqian.chen@utsa.edu}
\affiliation{
    \institution{The University of Texas at San Antonio}
    \country{USA}
}

\author{Ahmad-Reza Sadeghi}
\email{ahmad.sadeghi@trust.tu-darmstadt.de}
\affiliation{%
  \institution{Technical University of Darmstadt}
  \country{Germany}
}

\begin{abstract}
%\boldmath
Federated learning (FL) is a distributed learning process that uses a trusted aggregation server to allow multiple parties (or clients) to collaboratively train a machine learning model without having them share their private data. Recent research, however, has demonstrated the effectiveness of inference and poisoning attacks on FL. Mitigating both attacks simultaneously is very challenging. \Sota solutions have proposed the use of poisoning defenses with Secure Multi-Party Computation (SMPC) and/or Differential Privacy (DP). However, these techniques are not efficient and fail to address the malicious intent behind the attacks, i.e., adversaries (curious servers and/or compromised clients) seek to exploit a system for monetization purposes. To overcome these limitations, we present a ledger-based FL framework known as \ourname that allows making parties accountable for their behavior and achieve reasonable efficiency for mitigating inference and poisoning attacks. Our solution leverages crypto-currency to increase party accountability by penalizing malicious behavior and rewarding benign conduct. We conduct an extensive evaluation on four public datasets: Reddit, MNIST, Fashion-MNIST, and CIFAR-10. Our experimental results demonstrate that (1) \ourname provides strong privacy guarantees for model updates without sacrificing model utility; (2) \ourname can successfully mitigate different poisoning attacks without degrading the performance of the global model; and (3) \ourname offers unique reward mechanisms to promote benign behavior during model training and/or model aggregation.

\end{abstract}


\keywords{\vspace{-0.035cm}blockchain, federated learning, homomorphic encryption, security and privacy}

\maketitle
\footnotetext[1]{Affiliated with The University of Texas at San Antonio during the research and preparation of this paper.}\setcounter{footnote}{1}
\footnotetext[2]{Affiliated with Technical University of Darmstadt during the research and preparation of this paper.}\setcounter{footnote}{2}

\section{Introduction}
In recent times Machine Learning (ML) has gained high popularity and it is used for an increasing number of applications. However, the need to collect a large amount of data to train ML models raises security and privacy concerns in applications where sensitive data (i.e., text messages typed on mobile phones, personal medical information) is constantly stored and manipulated. Federated Learning (FL) allows multiple parties holding private data to collaboratively train ML models. Rather than collecting all data on a central server, each party (client) trains a model locally (local model) and only shares its parameters with the coordinating server that aggregates the parameters from all individual clients. Afterward, this aggregation server distributes the aggregated model (global model) back to the clients for further training rounds. 
 
This privacy feature, in combination with the performance gained by outsourcing the training process from one server to multiple clients, made FL the ideal training framework for different real-world applications, e.g., word suggestions in the mobile keyboard GBoard~\cite{mcmahan2017googleGboard}, brain tumor segmentation~\cite{sheller2018multi}, risk detection on mobile devices~\cite{fereidooni2022fedcri}, or identification of malware infected devices~\cite{nguyen2019diot}.

However, recent work challenges the security and privacy of FL, raising concerns about its practical applicability. For example, it was recently demonstrated that, given the model's parameters or predictions, model inference attacks could extract information about the training data from a model. Therefore by exploiting the model's memorization capabilities (or over-fitting), it is possible to reconstruct samples from the training data or determine if a certain sample was used for training~\cite{liu2022threats, shokri2017membership}, negating the privacy gains of FL.
The most prominent example of this attack is the Membership Inference Attack~\cite{shokri2017membership}. FL is particularly vulnerable to inference attacks executed by a curious server, as it has access to the local models before aggregation. {While the aggregation anonymizes the individual clients' contributions and makes inference attacks significantly harder~\cite{fereidooni2021safelearn}, the server's access to individual local models poses a significant threat to the clients' privacy and raises concerns for applications with privacy-sensitive data~\cite{sheller2018multi}. To mitigate inference attacks, \sota defenses use one of the following approaches: Secure Multi-party Computation (SMPC)~\cite{ryffel2018generic,fereidooni2021safelearn,acar2017achieving}, Differential Privacy~(DP)~\cite{mcmahan2017learning,mondal2022beas}, or Multi-Key Homomorphic Encryption (MKHE) \cite{sav2020poseidon}. 

In terms of security, recent work~\cite{bagdasaryan2020backdoor, shen2016auror} has demonstrated the high vulnerability of FL systems against Poisoning Attacks}. In this type of attack, adversaries leverage the distributed property of FL to take control of one or more training clients. Malicious clients can alter their behavior and skew the convergence of the global model. Poisoning attacks can be divided into two categories: untargeted and targeted attacks. Untargeted attacks aim to degrade the utility of the global model~\cite{kairouz2021advances}. In targeted (or backdoor) attacks \cite{bagdasaryan2020backdoor, xie2019dba}, however, an adversary guides the global model to a well-defined outcome. One example of an effective backdoor attack is where a malicious model is carefully trained to maintain a high accuracy for the main task but triggers backdoor behavior if specific patterns are detected during inference. Examples for such backdoor behavior include, e.g., injecting advertisement into an FL-based word suggestion system or creating a backdoor that makes an FL-based network intrusion detection system fail to detect network traffic of certain malware. A major threat of these attacks is that models are still black boxes. Thus, in practice, it is still an unsolved problem to determine if a model contains a hidden backdoor. The unique behavior makes backdoor attacks more important compared to other poisoning attacks. \Sota defenses aim to minimize the threat of poisoning attacks using techniques such as model filtering \cite{shen2016auror,blanchard2017machine} to detect and exclude poisoned models from aggregation, and/or model clipping \cite{rieger2022deepsight,nguyen2022flame} to limit poisoned updates' impact. These approaches, however, are limited by the underlying assumptions imposed by SMPC, e.g., availability during computations, and more importantly, they fail to address the motivation behind client misbehavior, i.e., accountability. Without crediting the contributions of individual clients, a malicious client may continuously try to poison the model until it is successful in one round. Further, mitigating poisoning and inference attacks at the same time is a complex task and existing approaches~\cite{nguyen2022flame,khazbak2020mlguard} are not efficient.

To overcome the limitations of existing solutions, we tackle the following questions: i) how to achieve a privacy-preserving aggregation framework that penalizes malicious intent during the aggregation process, ii) how to discriminate poisoned from benign updates to dynamically \mbox{reward or punish client's behavioral patterns.} \\
\textbf{Goals and Contributions.} In this paper, we present the design, implementation, and evaluation of \ourname, a fully-decentralized crypto-system that provides resiliency to inference and poisoning attacks. \ourname is a 3-layer blockchain FL framework powered by smart contracts, where each layer operates specific components, i.e., training clients (client layer), smart contracts (computation layer), and ledger (data layer). Our primary motivation to use smart contracts is to provide a decentralized and immutable environment to protect the security and privacy of models. By using smart contracts, we achieve reasonable efficiency and are able to mitigate poisoning and inference attacks. \ourname leverages blockchain's decentralization to yield high computation availability and ledger immutability (i.e., committed data cannot be changed) to prevent data alterations that could lead to unexpected results such as inaccurate model filtering or incorrect distributions of rewards.

To address the first question and perform privacy-preserving aggregation, we have to consider the following factors: private computation framework and aggregator compensation. To design a privacy-preserving computation platform, \ourname introduces the concept of Blockchain Two Contract Computation (BT2C). BT2C is defined as a semi-honest relationship\footnote{The semi-honest setting is a well-established security model that dictates how involved parties must adhere to the pre-established protocol} between two smart contracts using Homomorphic Encryption (HE), in particular, we rely on the CKKS\footnote{Note that we use the term HE to refer to CKKS in the rest of this paper.} encryption scheme~\cite{cheon2017homomorphic}. 

Compared to SMPC and MKHE, BT2C is a decentralized crypto-system based on HE that leverages the blockchain ledger to improve trust among smart contracts, where one contract (Defender) acts as a decryption service and the other contract (Gateway) acts as a computation hub. For our implementation, we develop a \textit{secure decryption} method that includes a compensation algorithm to evaluate a reward for the aggregation service based on its behavior. Our approach operates as a semi-honest cryptographic service such that the Gateway contract receives encrypted models from training clients and performs computations; and the Defender contract evaluates model characteristics (e.g., cosine distance) and provides incentives (i.e., crypto-currency, tokens) for benign behavior. 

To address the second question and discriminate poisoned models, we first separate it into the following components: poisoning detection and client compensation. To implement the poisoning detection, \ourname calculates the cosine distances between local and global models, and utilizes the Gaussian Kernel Density Estimation (G-KDE) function to divide them into different clusters. Here, a cluster is identified by the location of local minimums\footnote{A local minimum is a point on the associated function (e.g., G-KDE) whose value is less than every other point in its vicinity.}. This information is leveraged as a breaking point to separate the distances into different clusters. After benign and malicious clusters have been correctly identified, \ourname implements a round-based client compensation algorithm to provide additional incentives to benign training behavior, and to penalize \mbox{those who attempt model poisoning.}

In summary, \ourname's contributions are threefold:

(1) \ourname offers strong privacy guarantees by operating models in cipher text using the proposed BT2C protocol. Our approach is shown to be resilient against white-box inference attacks with a probability of success of $\frac{1}{m!}$, where $m$ represents the number of ciphers generated per model (\sect~\ref{sec:eval-miaMitigation}). 

(2) \ourname mitigates poisoning attacks using the proposed G-KDE clustering method to analyze the distribution of cosine distances and remove poisoned models. Our extensive evaluation on four public datasets (i.e., Reddit, MNIST, Fashion-MNIST, and CIFAR-10) indicates that \ourname is resilient against untargeted and targeted poisoning attacks (\sect~\ref{sec:eval-backdoorMitigation}).

(3) \ourname relies on our proposed aggregation and training compensation algorithms to offer incentives to benign aggregation services and benign training clients. Our results indicate that the proposed compensation algorithms automatically adjust the rewards to deter malicious intent from the training process (\sect\ref{sec:reward_system}).

\section{Requirements and Challenges}
This section presents the security and privacy requirements that \ourname fulfills and the challenges to be tackled in achieving them.  

\subsection{Privacy for FL}
During model submission, clients upload trained models to the aggregation server such that the server generates a new global model  (see App.~\ref{app:fl_background} for details on the FL process). At this point, the server has complete access to each model (e.g., model weights, structure and hyperparameters), which increases the threat of white-box inference attacks. To mitigate the attack, the defender has to satisfy the following requirements: \\
\textbf{P1: Utility Retention.} The defense must provide resiliency against inference attacks that are executed by curious servers % MIAs
while maintaining the utility of the model, i.e., main task accuracy (MA) remains the same with or without defense. Therefore, the performance of the new global model must not be compromised with the increase of privacy levels. \\
\textbf{P2: Computation Availability.} The defense must remain available to process and analyze encrypted models\footnote{An encrypted model is a collection of ciphers that represent encrypted weights.}. Therefore, every model computation shall not fail due to limited resource availability.

To the best of our knowledge, existing solutions for inference attacks that also preserve model utility rely on frail computation infrastructures, e.g., SMPC \cite{fereidooni2021safelearn, nguyen2022flame} or MKHE \cite{sav2020poseidon}. Thus, they suffer from the requirement of high availability of the system's components to use privacy-preserving computations (SMPC and MKHE) \cite{sav2020poseidon} and also from a high computation complexity to detect poisoned models when using privacy-preserving computations. Our scheme combines blockchain (see App. \ref{app:block_background}) and HE, in particular, the scheme of Cheon-Kim-Kim-Song (CKKS)~\cite{cheon2017homomorphic} (see App.~\ref{app:he_background}), to introduce a unique privacy-preserving computation framework, overcoming current limitations.
However, the use of blockchain brings additional concerns. Thus, \ourname addresses the following challenges: \\
\textbf{C1:} How to leverage blockchain to improve trust between computation parties.  \\
\textbf{C2:} How to effectively combine HE and blockchain to limit the ledger's transparency effect and increase privacy to model updates.

\subsection{Security for FL}
FL is a distributed learning approach that allows numerous clients to participate in the training process through model submissions. An adversary who controls a fraction of the clients can then use their influence to poison the new global model. To mitigate poisoning attacks, the defender has to fulfill the additional requirements: \\
\textbf{S1: Effective Poisoning Mitigation.} The defense must detect poisoning attempts, e.g., untargeted and targeted attacks, minimize their impact on the global model, and preserve model utility. For example, for targeted (backdoor) attacks, a defense should maintain the backdoor accuracy (BA) at the same level as without the attack. In addition, similar to \textbf{P1}, the defense must not negatively affect the training process, e.g., decrease MA by removal of benign models. \\
\textbf{S2: Autonomous Behavior.} The defense must be flexible to adjust automatically to different strategies without manual configuration.

Like existing solutions, \ourname leverages the cosine distance between local models and the global model to cluster their scores dynamically. Our approach, however, leverages this information to apply a deterrent to malicious clients, which adds another layer of security. Therefore, \ourname addresses the additional challenges: \\
\textbf{C3:} How to solve the dilemma of preventing the server from analyzing the local models against inference attacks while the server has to inspect the local models to detect/mitigate poisoned models.\\
\textbf{C4:} How to discriminate poisoned models s.t. malicious clients can be correctly identified to receive disciplinary actions.\\
\textbf{C5:} How to credit the clients over multiple training rounds to make malicious clients accountable for their attacks.

\section{Adversary Model and Assumptions}
\label{adversary}
In this section, we describe the threat model and assumptions used for the rest of the paper. We highlight the adversary's capabilities and main objectives. 
\subsection{Privacy Threat}
\label{sect:adversary-privacy}
Classic FL implementations rely on an aggregation server to compute new global models every training round (see App.~\ref{app:fl_background}). However, a malicious aggregation server can extract private information from each of the local models, thus, raising privacy concerns. \\
\textbf{White-box Inference Attack Goal.} Aligned with previous research \cite{nguyen2022flame, fereidooni2021safelearn}, the \textit{honest-but-curious} aggregator instantiates the attack on local models $W_i$ before aggregation. In other words, an adversary \adversaryPrivacy is aware of any process happening in the aggregator, but remains \textit{honest}, i.e., continues to perform the aggregator's benign tasks, to avoid detection. However, \adversaryPrivacy is also \textit{curious}, having the ability to infer private information about the training data $D_i$ while processing $W_i$. Formally, \adversaryPrivacy leverages $W_i$ to learn if a given input $x$ was used as part of $D_i$, allowing \adversaryPrivacy to extract sensitive information from every local model. Aligned with previous work~\cite{fereidooni2021safelearn,andreina2020baffle,khazbak2020mlguard}, we focus on inference attacks on the local models, as the aggregation anonymizes the individual contributions.\\
\textbf{\adversaryPrivacy Capabilities.} We assume \adversaryPrivacy is in full control of the aggregation server s.t. \adversaryPrivacy has access to every local model submitted by clients. We also assume \adversaryPrivacy cannot compromise clients directly or affect any of the training processes. 

\subsection{Security Threat}\label{adversary_security}
Multiple clients are selected to improve model accuracy. This collaboration allows one or more clients to conduct malicious activities in any training round. \\
\textbf{Targeted Poisoning Attack Goals.} In a targeted poisoning attack, the adversary \adversaryPoisoning has the following goals: poisoning injection and defense evasion. For poisoning injection, \adversaryPoisoning manipulates local model $W_i$ to produce poisoned local model $W'_i$. $W'_i$ is then used to alter the behavior of global model $G_t$. In \sota targeted poisoning attacks (also called backdoor attacks), \adversaryPoisoning guides poisoned global model $G'_t$ to behave normally all the time except when a specific set of conditions or triggers are present in the input. To achieve its secondary goal, \adversaryPoisoning manipulates $W'_i$ s.t. it remains as close as possible to $W_i$, e.g., adapting the loss function \cite{bagdasaryan2020backdoor}. \\
\textbf{\adversaryPoisoning Capabilities.} Similar to recent studies \cite{bagdasaryan2020backdoor,andreina2020baffle,rieger2024crowdguard,kumari2023baybfed}, we assume \adversaryPoisoning maliciously controls $f$ compromised clients, which should be less than half of the total number of clients $n$ ($f < \frac{n}{2}$).  We also assume \adversaryPoisoning cannot observe benign clients' local data or their submitted local updates. To introduce a backdoor into the global model, \adversaryPoisoning can launch a combination between data poisoning \cite{shen2016auror} and model manipulation attacks\cite{bagdasaryan2020backdoor}. Data poisoning is when \adversaryPoisoning adds \textit{poisoned} data to the existing training sets during model training, e.g., for an image classification task, \adversaryPoisoning can poison an image by drawing a shape into a specific corner. This attack allows \adversaryPoisoning to change model predictions to its desired outputs every time a trigger is identified during inference. In contrast, the model manipulation attack lets \adversaryPoisoning control the training algorithm to alter the convergence point of a model. This attack can be implemented through model scaling, modifying the loss function and/or adapting dedicated hyperparameters.

\subsection{Assumptions}\label{assumptions}
\ourname provides numerous security and privacy benefits under the following assumptions. \\
\textbf{A1: Consensus Protocol is not Compromised.} Since blockchain is the underlying platform to exchange information and execute smart contracts, we assume the consensus process to be not compromised.\\
\textbf{A2: Non-colluding Servers (Smart Contracts).} During the manipulation of encrypted data, deployed smart contracts engage in a semi-honest relationship to enable a privacy-preserving aggregation infrastructure. Therefore, to preserve privacy guarantees, we assume an adversary cannot control both contracts and their storage components, simultaneously. \\
\textbf{A3: Clients Perform Encryption.} Training clients affiliated to \ourname are assumed to have sufficient computational resources to perform encryption.

\section{Design}
This section first provides a high-level overview of \ourname and then describes its components in detail.

\subsection{High-level Overview}\label{high_level}
\begin{figure}[!t]
    \centering
    \includegraphics[width=0.8\columnwidth]{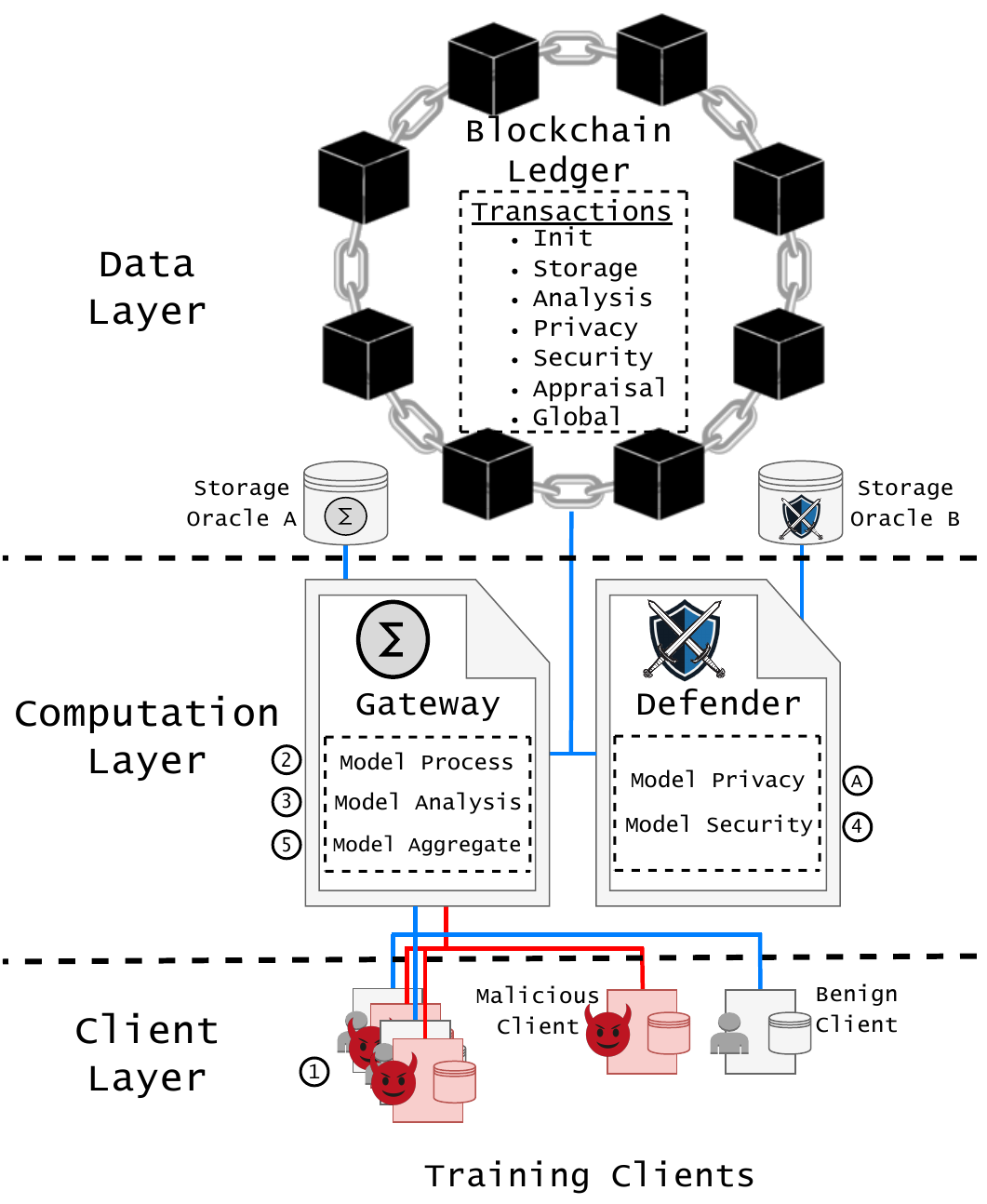}
    \caption{\ourname System Overview. Annotated steps illustrate the operation of \ourname during a training round $t$.}
    \label{fig:overview}
\end{figure}

\ourname is designed to fulfill privacy requirements (\textbf{P1}, \textbf{P2}) and security requirements (\textbf{S1}, \textbf{S2}). This is achieved by a layered framework that we will detail below. To detect poisoned models and satisfy \textbf{S1} and \textbf{S2}, \ourname uses a Gaussian Kernel Density Estimation (G-KDE) function to partition the received model updates into distinct clusters based on their pairwise distance. Since the cosine determines the angle, it reveals the changes that were applied to the local model. Compared to other metrics such as the Euclidean distance, it is more stable against manipulations. To satisfy \textbf{P1} and \textbf{S1} \ourname uses HE to encrypt models and perform privacy-preserving computations, i.e., private aggregation and/or private distance between models. In addition, \ourname leverages blockchain, in particular, smart contracts to meet \textbf{P2} and \textbf{S2}. Informally, \ourname is a 3-layer blockchain framework regulated by smart contracts that provides FL services to train models on arbitrary learning tasks, e.g., image classification, and/or word prediction. For every new learning task, a session reward is set by the owners of the task s.t. interested parties (clients and/or contracts) who join can be rewarded for their benign efforts. In other words, \ourname operates a crediting system to encourage participants to avoid malicious attempts to break the system, e.g., white-box inference or poisoning attacks. To manage every reward, \ourname registers training clients by generating unique cryptographic identities via its Membership Service Provider (MSP).\footnote{In Fabric, an MSP provides verifiable identities to members of the blockchain network.} Fig.~\ref{fig:overview} presents the different layers of \ourname.\\
\textbf{Client Layer.} This is the base layer of \ourname, and it is where training clients reside. As discussed in \sect\ref{adversary_security} some of the clients can be controlled by the poisoning attacker \adversaryPoisoning.\\
\textbf{Computation Layer.} This layer illustrates the logical components that enable \ourname to operate autonomously. It is formed by two smart contracts, namely Gateway and Defender contracts. The Gateway contract acts as the access gateway where clients submit their local models. Its core functions, e.g., model process, model analysis and model aggregate, provide privacy-preserving computations for encrypted models, addressing C1. The Defender contract, on the other hand, provides support to the Gateway in the form of security and privacy mechanisms, e.g., model privacy and model security, thus, allowing \ourname to defend against multiple threats. We define this blockchain infrastructure as Blockchain Two Contract Computation (BT2C). The goal of our BT2C implementation is to enable HE-based computations and secure decryption functions tailored to provide privacy-preserving FL (as defined in \sect\ref{sect:adversary-privacy}). Further details about the internal methods of both smart contracts can be found in Steps 2--5 of \sect\ref{design}. \\
\textbf{Data Layer.} The following layer represents the storage components of \ourname which constitutes two storage oracles, namely A and B, and a blockchain ledger. The storage oracles (i.e., external databases) are used to manage encrypted models and decryption keys for Gateway and Defender contracts, respectively, and the blockchain ledger stores information about \ourname (e.g., model information, session information) using the following transactions types (TT1 -- TT7). \\
\emph{Init Transactions (TT1)} are generated for each learning task to determine the owner of the task, the encryption keys to be used (i.e., HE public key $P_k$), the number of rounds $T$ required, and the reward amount $R$ for the full training session.\\
\emph{Storage Transactions (TT2)} are generated for every encrypted local model $W^*_i$ submitted by K clients to save the client ID (e.g., wallet address to receive payments), model ID, and the encrypted offset value $\delta^*_i$, where $i \in [1,K]$. The offset $\delta$ is a random value generated based on the standard deviation of local model $W_i$ that is injected before model encryption to further obfuscate $W^*_i$. Further information about the use of $\delta$ is found in Step 1 of \sect\ref{design}.\\
\emph{Analysis Transactions (TT3)} are created for every TT2 to compute the cosine distance between $W^*_i$ and the current encrypted global model $G^*_t$, where $t \in [1,T]$. TT3 is used to store model ID and its respective score $c_i$, i.e., cosine distance. \\
\emph{Privacy Transactions (TT4)} are generated when a malicious contract (i.e., Gateway) is attempting to break the privacy of $W^*_i$. TT4 includes the computed aggregation reward $R_C$ for a given training session. Further information for $R_C$ is found in Step A of \sect\ref{design}.\\
\emph{Security Transactions (TT5)} are created after \ourname evaluates every $c_i$ from TT3 to classify clients into two categories: benign or malicious. This is represented in TT5 as a list of benign IDs and a list of malicious IDs.\\
\emph{Appraisal Transactions (TT6)} are determined after TT5 to calculate the round reward $R_\tau$ for benign clients. Additional details for $R_\tau$ can be found in Step 4 of \sect\ref{design}. \\
\emph{Global Transactions (TT7)} are determined after model aggregation has occurred. TT7 includes the global ID, the decrypted weights of the new global model $G_{t+1}$, and the corresponding encrypted global weights $G^*_{t+1}$. 

\subsection{\ourname Details}\label{design}
To initialize \ourname, the smart contracts seen in the computation layer are deployed and initialized to the blockchain network. The initialization process for the smart contracts is completed after generating TT1. To avoid the possibility of data modification (or forks) at run-time, we refer back to assumption \textbf{A1}. Similarly, each smart contract takes into consideration assumption \textbf{A1} to protect the integrity of the contracts before/after deployment into the blockchain, and assumption \textbf{A2} to prevent an adversary from gaining full control over the system. Finally, to achieve privacy-preserving computations (as defined in \sect\ref{sect:adversary-privacy}), clients are bound to assumption \textbf{A3} to protect the privacy of local updates. 
\begin{algorithm}[b]
\caption{BT2C -- Private Cosine Distance}\label{alg:distance}
    \SetAlgoLined
    \SetKwInOut{Input}{Input}
    \DontPrintSemicolon

    \Input{$\delta^*$ $\triangleleft$ encrypted offset \newline
            $G^*$ $\triangleleft$ encrypted global model \newline
            $W^*$ $\triangleleft$ encrypted local model}
    $Z_D \leftarrow$ PrivateDotProduct($G^* + \delta^*$, $W^*$)
    
    $X_D \leftarrow$ SecureDecryption($Z_D$) $\triangleleft$ defender function
    
    $Z_G \leftarrow$ PrivateMagnitudeSquared($G^* + \delta^*$)
    
    $X_G \leftarrow$ SecureDecryption($Z_G$)
    
    $Z_L \leftarrow$ PrivateMagnitudeSquared($W^*$)
    
    $X_L \leftarrow$ SecureDecryption($Z_L$)
    
    $c \leftarrow 1-\frac{\sum_{i=1}^{n} X_{D_i}}{\sqrt{\sum_{i=1}^{n} X_{G_i}} * \sqrt{\sum_{i=1}^{n} X_{L_i}}}$ 
    
    UpdateScoreToLedger($c$) $\triangleleft$ new TT3
    
\end{algorithm}

The annotated steps seen in Fig.~\ref{fig:overview} illustrate the learning process of \ourname during a training round $t$. Here, we separate the learning process into 6 steps, i.e., 5 main steps (Step 1 -- Step 5) and 1 intermediary step (Step A). After multiple clients have joined a learning task, they first download the previous global model $G_{t-1}$ and the corresponding encryption key $P_k$ from TT1 at the Gateway contract. Note that for every $P_k$, a corresponding secret key $S_k$ is generated and maintained by the Defender contract. Furthermore, every $P_k$ includes an encryption context, which is provided to the clients. The encryption context contains the degree of the polynomial (\polydeg) used to generate $P_k$ and $S_k$. This value determines the size of $P_k$ and the size of produced ciphers in terms of bytes. Clients may continue to use the same $P_k$ unless a new public key is required by the system. \\
\textbf{Model Encryption (Step 1)}. Each client $i$ starts to train the model using local data $D_i$ for a predefined number of epochs. After training, clients generate and inject an offset constant $\delta_i$ such that $W'_i = W_i + \delta_i$. More specifically, $\delta_i$ is generated from the multiplication of two random elements: the model's standard deviation $\sigma_{W_i}$ after training, and a scaling factor $f_s \in [-100,100]$ s.t. $f_s \neq 0$. Note that $f_s$ is bounded to $[-100, 100]$ to avoid exceedingly large numbers (positive or negative) as model weights. This is primarily because we are interested in shifting (left or right) the distribution of $W_i$ using the inherently random properties of each local model. Contrary to DP, $\delta_i$ is recorded to be used in Step 3. The offset is applied to mask $W_i$ and obfuscate the model during private computations. At this stage, an attacker would require to brute force every $\delta_i$ in order to break the privacy of a single local update. Then clients start the encryption process of $W'_i$ and $\delta_i$. Once a client has generated its encrypted local model $W^*_i$ and encrypted offset $\delta^*_i$, the client proceeds to submit them to the Gateway contract for further analysis. By this, \ourname addresses \textbf{C2}.

Due to the limitations of HE, a client is required to first separate $W'_i$ into multiple chunks of data s.t. $W'_i = w'_1...w'_n$, where $n$ is the number of chunks per model. To calculate $n$, we first determine the capacity for every cipher $z$, i.e., the maximum amount of elements each cipher is able to contain. We define the capacity to be $\polydeg \mathbin{/} 2$, e.g., a \polydeg of 2048 yields a capacity of 1024 elements per cipher. Thus, the number of ciphers required to encrypt $W'_i$ is directly proportional to the number of trainable parameters given \polydeg. This is further illustrated by Eq.~\ref{eq:encrypt}:\\
\begin{equation}\label{eq:encrypt}
    z_1...z_n = \textnormal{Encrypt($w'_1...w'_n$, $P_k$), } n = \frac{\textnormal{len($W'_i$)}}{\polydeg / 2} + 1
\end{equation}

\begin{algorithm}[!t]
\caption{BT2C -- Secure Decryption}\label{alg:decryption}

    \SetAlgoLined
    \SetKwInOut{Input}{Input}
    \SetKwInOut{Output}{Output}
    \DontPrintSemicolon

    \Input{$z_1, \dots, z_m$ $\triangleleft$ computation ciphers}
    \Output{$X$ $\triangleleft$ array of decrypted numbers \newline
             $\rho$ $\triangleleft$ array of decrypted model chunks} 
     $\delta^*_1, \dots, \delta^*_K \leftarrow$ ReadOffsetFromLedger() $\triangleleft$ from TT2
    
     $S_k \leftarrow$ ReadKeyFromStorage() 
    
     $t \leftarrow 0.05$ $\triangleleft$ array variation tolerance
     
    \For{each cipher $i$ in [$1,m$]}{
        $\rho_i \leftarrow$ Decrypt($z_i$, $S_k$)
        
        $v \leftarrow \mid \frac{\max (\rho_i) - \min (\rho_i)}{\max (\rho_i)} \mid$ $\triangleleft$ compute variation
     
        \uIf{$v \leq t$}{
            $X_i \leftarrow$ Average($\rho_i$)
        }
        \uElseIf{$K > 1$}{
            \For{each offset $j$ in [$1,K$]}{
                $\delta_j \leftarrow$ Decrypt($\delta^*_j$, $S_k$)
            }
            
            $\rho_i \leftarrow \frac{\rho_i - \sum_{j=1}^{K} \delta_j}{K}$ $\triangleleft$ offset removal/injection
        
        }\Else{
            $R \leftarrow$ ReadRewardFromLedger() $\triangleleft$ from TT1

            $s \leftarrow$ CountSessionsFromLedger() $\triangleleft$ \# TT1

            $\phi \leftarrow$ CountAnomaliesFromLedger() $\triangleleft$ \# TT4

            $R_C$ $\leftarrow$ $0.1 * R * e^{-(\phi + 1) / s}$ $\triangleleft$ calculating reward 

            UpdateContractRewardToLedger($R_C$) $\triangleleft$ new TT4
                
            $\rho_i \leftarrow$ $\emptyset$ $\triangleleft$ empty set
            %\Return{} $\triangleleft$ terminate BT2C round
        }
         
    }
    
    \Return{$X$ or $p$} $\triangleleft$ output type dependent on process 
\end{algorithm}
For \ourname, we have determined a minimum \polydeg of 4096 is required to successfully compute the desired private functions. \\
\textbf{Model Process (Step 2)} is the initial function that receives every $W^*_i$ provided by the clients. In this step, the Gateway contract stores $W^*_i$ into storage oracle A to avoid public visibility to any other contract deployed in the network. The storing process saves the ciphers as encoded text into a single document. Every pair of $W^*_i$ and $\delta^*_i$ is used to generate and submit a new TT2 to the ledger. \\
\textbf{Model Analysis (Step 3)} uses the previously submitted TT2 to retrieve the encrypted model from storage and its corresponding encrypted offset. $\delta^*_i$ is used to offset the encrypted global model $G^*_{t-1}$, where $G^*_{t-1}$ can be easily downloaded from TT7 of the previous round. This process aligns encrypted models to compute an accurate cosine distance as given by Alg.~\ref{alg:distance}.

Formally, the private cosine distance function seen in Alg.~\ref{alg:distance} requires as inputs the encrypted global model $G^*$, the encrypted local model $W^*$ and the corresponding encrypted offset $\delta^*$. Its goal is to compute the cosine distance score $c$ between $G^*$ and $W^*$. To calculate the distance, the computation process is segmented into three BT2C rounds. This is to overcome the practical limitations of HE, e.g., inability to compute roots. The first round (lines 1--2) starts by computing the encrypted dot product $Z_D$ between $G^* + \delta^*$ and $W^*$, where $Z_D$ is a collection of ciphers $z_1,\dots,z_n$ that represent the encrypted value of the dot product operation. $Z_D$ is then delivered to Defender contract to perform \textit{secure decryption}. Note that $Z_D$ might be in any order to add randomness to the decryption process. This process returns $X_D$, a collection of numbers $x_1,\dots,x_n$ that represents the results of $\sum G \cdot W$. Similarly, the two remaining rounds (lines 3--4 and 5--6) are used to generate $X_G$ ($\sum G^2$) and $X_L$ ($\sum L^2$), respectively. To finalize the computation process, the values for all three rounds are combined to calculate $c$ and submitted to the ledger (TT3), as seen in line 7--8.\\
\textbf{Model Privacy (Step A)} relies on TT1, TT2 and previous TT4 to enable the \textit{secure decryption} function seen in Alg.~\ref{alg:decryption}. This function includes two unique operations: limitation of data decryption (lines 1--13) and reward adjustment (lines 15--20). The former analyzes the information in every cipher to either return a collection of numbers $X$ (see Step 3) or a collection of model chunks $\rho$ (see Step 5). The latter regulates the contract reward $R_C$ using the information stored in the ledger. The use of $R_C$ enables \ourname to incentivize benign aggregation behavior in the framework while penalizing malicious conduct such as attempting to access local models. Note that in \ourname, $R_C$ is set to be 10\% (max) of the session reward $R$ by default. 

Alg.~\ref{alg:decryption} requires as inputs only the computation ciphers $z_1,\dots,z_m$, where $m$ is the number of submitted ciphers in the BT2C round. Note that for secure decryption, $m$ is independent from the number of ciphers $n$ in an encrypted model such that $m \leq n$.\footnote{A practical example is when the aggregation contract (Gateway) divides its computations into multiple rounds s.t. $Z_m \in Z_n$ to prevent the Defender from potentially accessing full data.} Our approach returns decrypted data, which is represented by an array of numbers $X$ or an array of model chunks $\rho$. To initiate the decryption process and address \textbf{C3}, the Defender contract retrieves (lines 1--2) every encrypted offset $\delta^*_1,\dots,\delta^*_K$ from TT2, and the corresponding secret key $S_k$ from storage oracle B. The variation tolerance value in line 3 is set to $t=0.05$ (or 5\%) as it is required to discriminate summation operations, e.g., $\sum G \cdot W$, from model operations, e.g., model aggregation. More specifically, summation operations are determined by a low array variation, which indicates that all elements are the same or closely related\footnote{Array symmetry comes natively in HE to maintain optimum conditions during computations.}; and model operations are characterized with high variations as these are represented by distinct values within the decrypted array. After decryption (line 5), the data is analyzed w.r.t. the array variation factor $v$ (line 6), where $v$ is defined as the absolute percent difference between the maximum and minimum elements within $\rho_i$. At this step, if $v \leq t$, the elements inside $\rho_i$ are considered to be the result of a summation operation, thus, generating $X_i$ to represent its average as shown in line 8. Otherwise, the function proceeds to consider $\rho_i$ as a model operation, where $\rho_i$ is then adjusted by every $\delta_i$ and divided by the number of available models $K$ (from TT2) to complete the aggregation process $\sum_{i=1}^{K} \frac{W_i}{K}$ as defined by lines 9--13. 

\begin{algorithm}[!t]
\caption{Poisoning Defense}\label{alg:clustering}
    \SetAlgoLined
    \SetKwInOut{Input}{Input}
    \DontPrintSemicolon

    \Input{($c_i, \dots ,c_K$) $\triangleleft$ distance scores}
    
    $f \leftarrow 2000$ $\triangleleft$ resolution factor for smooth curves
    
    $(x_1, \dots ,x_f),(y_1, \dots , y_f) \leftarrow$ GaussianKDE($[c_i, \dots ,c_K]$, $f$) $\triangleleft$ compute gaussian kernel density estimation
    
    $(l_1, \dots ,l_N) \leftarrow$ LocalMinimums([$y_1, \dots ,y_f$]) $\triangleleft$ $l_n$ is the index of local minimum found in $y$
    
    $G \leftarrow \{[x_1,x_{l_1}], \dots ,[x_{l_{N-1}}, x_{l_N}], [x_{l_N},x_f]\}$ $\triangleleft$ group set based on local minimums
    
    $M \leftarrow N+1$ $\triangleleft$ maximum number of available groups
    
    \For{each group $m$ in $[1,M]$}{
        \For{each score $i$ in $[1,K]$}{
            \If{$c_i \in G_m$}{
                $g_m \leftarrow i$ $\triangleleft$ append model index $i$ to a group
            }
        }
    }
    
    UpdateGroupsToLedger($g$) $\triangleleft$ new TT5. $g_1$ is closest to $G_{t-1}$
    
    $R \leftarrow$ ReadRewardFromLedger() $\triangleleft$ from TT1
    
    $T \leftarrow$ ReadTotalNumberOfRoundsFromLedger() $\triangleleft$ from TT1
    
    $R_C \leftarrow$ ReadContractRewardFromLedger() $\triangleleft$ from TT4
    
    $R_\tau \leftarrow$ $\frac{R - R_C}{T * \textnormal{len($g_1$)}} $ $\triangleleft$ training reward
    
    UpdateTrainingRewardToLedger($R_\tau$) $\triangleleft$ new TT6

\end{algorithm}

Lines 15--20, however, are used to assess $K$ since aggregation must be performed with a minimum of $K=2$ models. If $K \leq 1$, it adjusts $R_C$ to penalize the contract for attempting to bridge the confidentiality of the first local model as this would not be obfuscated by multiple offset constants. This subroutine is performed by gathering the current session reward $R$ and the number of successful training sessions $s$ from TT1 (lines 15--16), and the number of privacy anomalies $\phi$ previously registered from TT4 (line 17). Note that $\phi$ increases every time $K \leq 1$. To finalize the penalization process, the new $R_C$ is generated and added to the ledger (TT4) as seen in lines 18--19, and $\rho$ is set to empty to \mbox{avoid leaking information (line 20).}\\ 
\textbf{Model Security (Step 4)} collects the cosine distance scores $c_i$ from TT3, applies the proposed clustering technique to filter poisoned models, and determines the client reward $R_\tau$ to promote benign training behavior. To remove poisoned updates and address \textbf{C4}, our clustering method uses the Gaussian Kernel Density Estimation (G-KDE) function to identify the number of data distributions within the distance scores $c_i$, and selects models according to their assigned distribution. If models are determined to be malicious, they are removed from the aggregation process and penalized by receiving a reward of 0 for that round. These leftover rewards are divided evenly among other clients, thus, addressing \textbf{C5}. Additional information related to the use of G-KDE to generate clusters is discussed in App.~\ref{app:intuition}.

The poisoning defense is presented in Alg.~\ref{alg:clustering}. The defense requires distance scores $c_1,\dots,c_K$ as input to generate model clusters (or groups). To produce an accurate representation of the distribution between scores, we set a resolution factor $f$ of 2000 (line 1), where $f$ denotes the number of data points used to fit the G-KDE function. Note that we empirically found that $f=2000$ provides the necessary resolution to find local minimums. In line 2, $c_1,\dots,c_K$ and $f$ are used to compute G-KDE, thus, generating 2000 ($x,y$) data points, where $x$ is bounded between $\min (c_1, \dots ,c_K)$ and $\max (c_1, \dots, c_K)$, and $y$ is the density estimation obtained from the process. Density values $y_1, \dots, y_f$ are used in line 3 to calculate the location or index of every local minimum $l_1, \dots ,l_N$ found in the distribution of scores. These locations are used to generate a group set $G$ that contains $N+1$ ($M$) groups of models (line 4). At this stage, each score is allocated inside specific groups $g$ to separate benign models from malicious (lines 6--12). These groups are committed to the ledger as part of TT5 in line 13. Finally, to calculate $R_\tau$, the process retrieves $R$ and the number of training rounds $T$ from TT1, the current $R_C$ from TT4, and the group closest to $G_{t-1}$ ($g_1$) are combined by the defense as seen in lines 14--17. The updated $R_\tau$ is \mbox{committed to the ledger as a new TT6 (line 18). }\\
\textbf{Model Aggregate (Step 5)} selects the models defined by TT5 from storage oracle A to compute a new global model $G_t$. The private aggregation (Alg.~\ref{alg:aggregation}) uses a single {BT2C computation round to create $G_t$.}

\begin{algorithm}[!t]
\caption{BT2C -- Private Aggregation}\label{alg:aggregation}
    \SetAlgoLined
    \SetKwInOut{Input}{Input}
    \DontPrintSemicolon
    
    \Input{$W^*_1, \dots, W^*_N$ $\triangleleft$ selected models}
    $Z \leftarrow W^*_1$ $\triangleleft$ encrypted base model
        
    \For{each update $i$ in [$2,N$]}{
        $Z \leftarrow$ Add($Z$,$W^*_i$)
    }
        
    $G_t \leftarrow$ SecureDecryption($Z$) $\triangleleft$ defender function
    
    $P_k \leftarrow$ ReadKeyFromLedger() $\triangleleft$ from TT1   
    
    $G^*_t \leftarrow$ Encrypt($G_t$, $P_k$)
    
    UpdateGlobalToLedger($G^*_t$, $G_t$) $\triangleleft$ new TT7
\end{algorithm}

Formally, Alg.~\ref{alg:aggregation} requires as inputs every selected local model $W^*_1, \dots, W^*_N$, where $N$ is the number of admitted models selected in Step 4. Models are simply added (lines 1-4) into a single encrypted model $Z$, where $Z=(z_1,\dots,z_n)$. In line 5, $Z$ is submitted to Defender contract to complete the aggregation process, which returns a collection of arrays denoted as $G_t$. The new model is then encrypted (lines 6--7) using the available $P_k$ to produce the new encrypted global model $G^*_t$. $G_t$ and $G^*_t$ are compiled into a new TT7 and committed into the ledger (line 8) to prepare \ourname for the next training round.

\section{Experimental Setup}
The following sections illustrate our testbed, and describe the datasets and models used during evaluation. Note that a detailed list of evaluation metrics is provided in App.~\ref{app:metrics}. \\
\textbf{Experimental Testbed.} We simulate a generic blockchain using Hyperledger Fabric (HLF) to illustrate the practicality of our approach for other blockchain implementations. For experimental setup, we abstract away the complexities introduced by the consensus protocol, and instead focus on the computational entanglements added by \ourname. This is because \ourname relies solely on smart contracts rather than the underlying blockchain platform. To instantiate the simulation environment, we deploy docker containers on a Windows PC with Intel Core i7-9750H and 32 GB RAM. The blockchain test network is formed by a single ordering node operated by single organization with two peers transacting under a single communication channel.
\begin{table}[t]
\caption{Dataset description for different learning tasks.}
\label{tab:dataset}
\centering
    \begin{tabular}{c|ccc|c}
    \hline
    \textbf{Application} & \multicolumn{3}{c|}{\textbf{IC}} & \textbf{WP} \\
    \hline
    \textbf{Datasets} & MNIST & Fashion & CIFAR-10 & Reddit \\
    \textbf{\#Records} & 70K & 70K & 60K & 20.6M \\ 
    \textbf{Model} & CNN & CNN & $\textnormal{ConvMixer}_{256/3}$ & LSTM \\
    \textbf{\#params} & $\sim23$K & $\sim29$K & $\sim234$K & $\sim20$M \\
    \textbf{\#ciphers} & 12 & 15 & 115 & $\sim10.1$K \\ 

    \end{tabular}%
    
\end{table} 

To fit multiple encrypted models within a single block, we increase the block size to 100MB. Note that this is 100 times larger than a Bitcoin block (1MB) \cite{nakamoto2019bitcoin}. We implement the gateway and defender contracts (chaincodes) using NodeJS and the Node-SEAL library. Node-SEAL
 is a NodeJS wrapper library used to interface with Microsoft SEAL \cite{chen2017simple}, an efficient and open-source HE library available in C++. Our HE setup uses a \polydeg of 4096 to encrypt local models. To evaluate models, we use Pytorch in an Ubuntu 20 server with 2 AMD EPYC 7302, 480 GB RAM, and 6 NVIDIA A100 (40 GB RAM each).\\ 
\textbf{Datasets and Models.} To assess \ourname, we use two popular FL applications: word prediction (WP) \cite{mcmahan2017communication,mcmahan2017learning}, and image classification (IC) \cite{sheller2018multi,chilimbi2014project,li2023flairs}. Note that every model used for evaluation has been pre-trained to reach an acceptable accuracy level. Tab.~\ref{tab:dataset} describes the dataset types (Dataset), the rounded number of records per dataset (\#Records), the AI models used for training (Model), the number of trainable parameters (\#params) and the number of ciphers (\#ciphers) found per model.  We use smaller models with fewer trainable parameters for IC datasets, compared with WP, to evaluate how model complexity impacts \ourname.\\
\emph{Word Prediction (WP).} We use the Reddit dataset as example of WP for Natural Language Processing (NLP) applications, e.g., the real-world FL application G-Board~\cite{mcmahan2017googleGboard}. The dataset contains over 20M records of reddit users' posts from November 2017. Following previous work \cite{rieger2022deepsight,bagdasaryan2020backdoor} we use a 2-layer LSTM model.\\
\emph{Image Classification (IC).} We selected three popular IC datasets of different image complexity: MNIST, Fashion-MNIST (or Fashion for short) and CIFAR-10. They all consist of 10 evenly divided categories, where MNIST contains 70K handwritten digits, Fashion has 70K images of articles of clothing (i.e., shoe, dress, shirt), and CIFAR-10 has 60K pictures of objects (i.e., frog, airplane, car). For MNIST and Fashion, we use a simple CNN model comprised of 1 and 2 CNN layers, respectively. We customized the ConvMixer model \cite{trockman2022patches} to train CIFAR-10 with a width of 256. \\
\textbf{Backdoor Attacks.} Aligned with earlier work~\cite{bagdasaryan2020backdoor,andreina2020baffle,nguyen2022flame} we use the constrain-and-scale attack of Bagdsaryan \etal~\cite{bagdasaryan2020backdoor}. Note that we focus on adaptive attacks, e.g., adversary adapts the loss function using the same metric as the defensive strategy. In other words, our adversary model leverages the \textit{cosine distance} in an attempt to evade our defense. For the Reddit dataset, the adversary aims to make the model predicting the word "delicious" after the trigger "pasta from astoria tastes"~\cite{bagdasaryan2020backdoor}. The CIFAR-10 backdoor shall make all cars in front of a striped background being classified as birds~\cite{bagdasaryan2020backdoor}. For MNIST and Fashion MNIST, the backdoor forces models to predict the number 0, and t-shirt/top, respectively, when the image trigger is detected. The trigger is simply a white rectangle located at the bottom left corner of each poisoned image.

\section{Experimental Results}
\label{results}
The following sections evaluate the privacy of \ourname under a naive setup, the security aspect of \ourname against poisoning attacks, and the behavior of the reward system in \ourname for aggregation services and clients. We also provide a run-time performance analysis of \ourname in App.~\ref{app:performance}, which is used to illustrate the increased complexity of our learning process.  

\begin{figure}[t]
    \centering    
    \includegraphics[width=0.85\columnwidth]{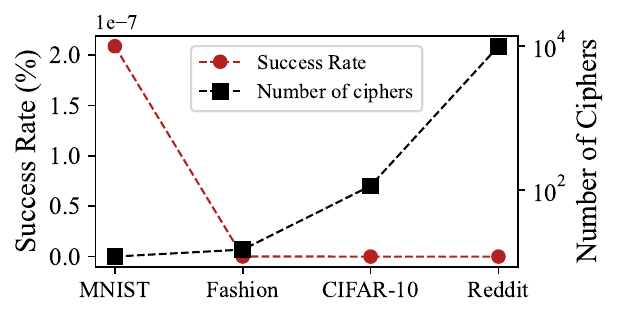}
    \caption{Probability of success for white-box inference attack w.r.t. model complexity.}
    \label{fig:decryption}
\end{figure}
\begin{table}[b]
\caption{Effectiveness of \ourname against multiple poisoning attacks in terms of Backdoor Accuracy \% (BA) and Main Task Accuracy \% (MA).}
\label{tab:stats}
\centering
    \begin{tabular}{c|c|c|c|c|c}
    \hline
   \multirow{2}{*}{\textbf{Poisoning Attack}} & \multirow{2}{*}{\textbf{Dataset}} & \multicolumn{2}{c|}{\textbf{No Defense}} & \multicolumn{2}{c}{\textbf{\ourname}} \\
    \cline{3-6}
    & & BA  & MA   & BA  & MA  \\
    \hline
    \multirow{5}{*}{Untargeted \cite{hossain2021desmp}} & Reddit & -- & 15.8 & -- & 22.7 \\
    & MNIST & -- & 91.5 & -- & 98.3 \\
    & Fashion & -- & 41.1 & -- & 90.0 \\
    & CIFAR-10 & -- & 28.9 & -- & 83.0 \\
    \hline
    \multirow{5}{*}{Constrain-and-Scale \cite{bagdasaryan2020backdoor}} & Reddit & 100 & 22.6 & 0.0 & 22.7 \\
    & MNIST & 98.0 & 87.7 & 0.4 & 98.3 \\
    & Fashion & 100.0 & 69.3 & 2.4 & 90.6 \\
    & CIFAR-10 & 100.0 & 66.1 & 0.0 & 83.8 \\
    \hline
    \multirow{5}{*}{DBA \cite{xie2019dba}} & Reddit & 100.0 & 22.6 & 0.0 & 22.7 \\
    & MNIST & 82.6 & 77.2 & 0.1 & 98.3 \\
    & Fashion & 99.7 & 36.7 & 1.0 & 98.3 \\
    & CIFAR-10 & 85.2 & 67.4 & 2.1 & 83.8 \\
    \end{tabular}
\end{table}
\subsection{White-box Inference Attack Resiliency}
\label{sec:eval-miaMitigation}
\textbf{Evaluation Baseline.} To evaluate the privacy of \ourname, we step outside \textbf{A2} to explore a limited collaboration between the Gateway and Defender contracts. In this scenario, the Gateway is in full control of the attacker and attempts aggregation when there is only one model in \ourname ($K=1$), disregarding its potential reward $R_C$. The Defender, however, is only partially compromised allowing the attacker to observe $\rho_i$ during secure decryption such that the attacker can reverse the offset from local model $W_1$. \\
\textbf{Effectiveness of \ourname.} We evaluate the effectiveness of the obfuscation techniques implemented in \ourname to prevent white-box inference attacks. To breach the privacy of $W_1$, a Defender requires to find the correct order of ciphers, since this is random for every BT2C computation round. We define such a brute force process to be equivalent to $m!$, where $m$ is the number of ciphers for the BT2C round. Therefore, the probability of an attacker finding the right combination of ciphers to generate the correct model $W_1$ is given by $1/m!$. Fig.~\ref{fig:decryption} illustrates the potential success rate ($1/m!$) of an attacker to extract $W_1$, and its contrast w.r.t. $m$ for every dataset. Here, we observe that the probability of success decreases from $2.08\mathrm{e}{-7}$\% to 0\% as the number of ciphers increases from 12 (MNIST model) to $\sim10.1K$ (Reddit model), i.e., large models provide better resiliency. Therefore, \ourname is $1/m!$ resilient to white-box inference attacks even during a limited collaboration outside \textbf{A2}.

\subsection{Poisoning Mitigation}
\label{sec:eval-backdoorMitigation}
\textbf{Evaluation Baseline.} We set PMR=0.5, non-IID=0.7, PDR=0.5 and $\alpha$=0.7 as baseline parameters for untargeted and targeted attacks (unless otherwise indicated). PMR (or Poisoned Model Rate) indicates the influence level of an attacker to the system, i.e., PMR of 0.5 denotes an attacker maliciously controls  50\% of training clients. Non-IID data (or non-Independent and Identically Distributed) represents the number of training samples dispersed to a client that belongs to a specific class within a pre-defined group, i.e., non-IID of 0.7 means that clients should use 70\% of training data from their given class while the rest 30\% is from the remaining classes. We follow the approach in \cite{fang2020local} to prepare each  dataset according to the number of output classes. PDR (or Poisoned Data Rate) determines the fraction of poisoned samples with respect to benign samples during training, i.e., PDR of 0.5 defines 50\% of training data from a target class are poisoned samples. A higher PDR increases the success rate of the attacks. Similarly, the regularization term $\alpha$, as defined by Bagdasaryan \etal \cite{bagdasaryan2020backdoor}, balances the loss function of client models aiming at limiting the distance between local and global models. A high value of $\alpha$ allows the attacker to increase its success rate at the cost of visibility. \\
\begin{table}[!t]
\caption{Comparison of \ourname and five state-of-art defenses' efficiency to mitigate backdoor. BA refers to  Backdoor Accuracy \%  and MA refers to Main Task Accuracy \%.}
\label{tab:defenses}
\centering
    \begin{tabular}{l|r|r|r|r|r|r|r|r}
    \hline
    \multirow{2}{*}{\textbf{Defenses}} & \multicolumn{2}{c|}{\textbf{Reddit}} & \multicolumn{2}{c|}{\textbf{MNIST}} & \multicolumn{2}{c|}{\textbf{Fashion}} & \multicolumn{2}{c}{\textbf{CIFAR-10}} \\
    \cline{2-9}
    & BA & MA & BA & MA & BA & MA & BA & MA \\
    \hline
    Benign Setting & 0.0 & 22.7 & 0.5 & 98.3 & 3.7 & 90.9 & 0 & 83.9 \\
    No Defense & 100.0 & 22.7 & 98.0 & 87.7 & 100.0 & 69.2 & 100.0 & 66.1 \\
    \hline
    Krum \cite{blanchard2017machine} & 100.0 & 22.6 & 0.6 & \textbf{98.3} & 2.8 & 90.1 & \textbf{0.0} & 83.0 \\
    FoolsGold \cite{fung2020limitations} & \textbf{0.0} & \textbf{22.7} & 0.5 & \textbf{98.3} & 3.0 & 90.7 & \textbf{0.0} & 83.6 \\
    Auror \cite{shen2016auror} & 100.0 & 22.5 & 0.5 & \textbf{98.3} & 2.5 & \textbf{90.9} & \textbf{0.0} & \textbf{83.9} \\
    AFA \cite{munoz2019byzantine} & 100.0 & 22.6 & 83.1 & 94.2 & 97.9 & 87.3 & 100.0 & 66.5 \\
    DP \cite{mcmahan2017learning} & 77.0 & 22.0 & 26.5 & 97.3 & 52.2 & 88.6 & 60.0 & 76.6 \\ 
    \hline
    \ourname & \textbf{0.0} & \textbf{22.7} & \textbf{0.4} & \textbf{98.3} & \textbf{2.4} & 90.6 & \textbf{0.0} & 83.8 \\
    \end{tabular}
\end{table}

\textbf{Effectiveness of \ourname.} We evaluate the resiliency of \ourname against different poisoning attacks such as untargeted poisoning~\cite{hossain2021desmp}, constrain-and-scale~\cite{bagdasaryan2020backdoor} and DBA~\cite{xie2019dba}. The experimental results illustrated in Tab.~\ref{tab:stats} show that \ourname successfully mitigates these attacks without sacrificing benign performance (MA). For untargeted poisoning, the adversary successfully degrades model performance when no defenses are in place, reaching as low as 15.8\% (22.7\% original) for Reddit, and 28.9\% (83.9\% original) for CIFAR-10. During constrain-and-scale attacks, the adversary is able to inject a backdoor into the model with almost 100\% accuracy. Similarly, for DBA, the backdoor is injected into the global model with 80+\%. These attacks, however, are not effective against \ourname as BA$\approx$0 for every evaluated dataset.\footnote{In some applications, misclassifications are counted in favor of the BA if MA$<100\%$. For this reason, the BA is greater than 0\%, e.g., 2.4\% for Fashion, although the aggregated model does not contain the backdoor.} Moreover, \ourname maintains or even increases MA. Note that for the rest of the evaluation, we focus on targeted (or backdoor) attacks since they are the most sophisticated type of poisoning attacks. \\
\textbf{Comparison to Existing Work.} Tab.~\ref{tab:defenses} compares the effectiveness of \ourname with five \sota defense approaches~\cite{blanchard2017machine,fung2020limitations,shen2016auror,munoz2019byzantine,mcmahan2017communication}. Notably, several defenses such as Krum~\cite{blanchard2017machine} cannot handle \nonIid scenarios. FoolsGold is the most resilient defense that mitigates backdoors for all four datasets as its BAs are very closed to 0, but still a little higher than \ourname's BA rates.
Similar to FoolsGold, the other four defenses' BA rates are  much higher than or equal to \ourname's while MA rates are much lower than or equal to ours. Auror~\cite{shen2016auror} works well for the image datasets, where the clients' local datasets overlap and show similar distributions but fails for the intrinsic non-IID Reddit dataset. Therefore, \ourname is shown to provide the most resilient defense to mitigate \sota backdoors. 
Appendix \ref{app:results_wp} and \ref{app:results_ic} provide further experiment results for WP and IC tasks respectively, showing \ournameGen performance for different PDRs, PMRs, further attack strategies, and IID rates.
\begin{figure}[t]
    \centering
    \includegraphics[width=0.975\columnwidth]{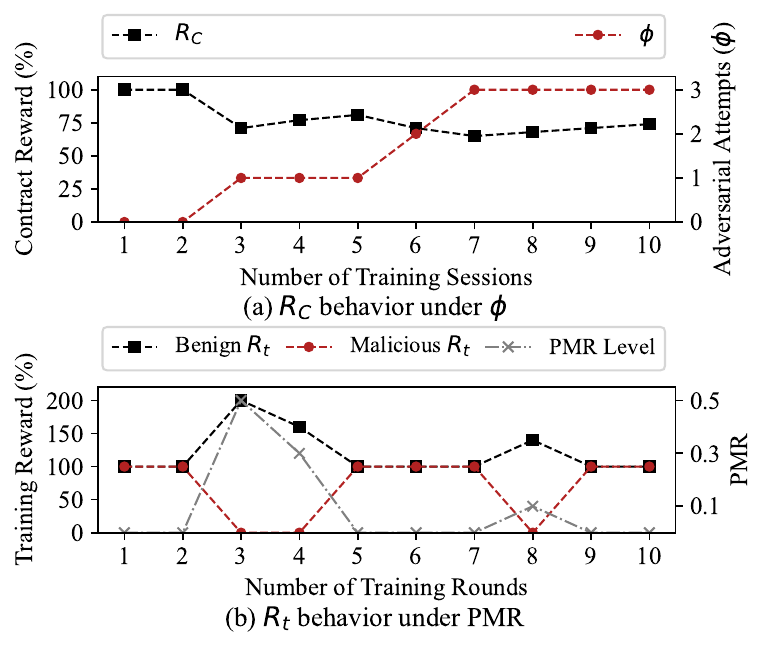}
    \caption{Behavior of rewards (a) $R_C$, (b) $R_t$ in \ourname.}
    \label{fig:rewards}
\end{figure}

\subsection{Reward Analysis}
\label{sec:reward_system}
The reward systems implemented in \ourname are an additional layer of security designed to discourage malicious actions during the learning process. However, a defensive strategy is only as good as its weakest component. In other words, \ourname's reward mechanisms are constrained by the efficiency of its white-box inference resiliency (see \sect~\ref{sec:eval-miaMitigation}) and its poisoning defense (see \sect~\ref{sec:eval-backdoorMitigation}). Thus, the following section investigates the rewarding mechanism behind \ourname. Fig.~\ref{fig:rewards} shows the behavior of the contract reward $R_C$ and the training reward $R_\tau$. \\
\textbf{Effect of $\phi$ in $R_C$.} We assume \ourname has received the first local model $W_1$ ($K=1$), and that the adversary has control over the Gateway contract. Fig.~\ref{fig:rewards}a illustrates how $R_C$ is adjusted by the Defender every time an attempt to access private models ($\phi$) is registered during \textit{secure decryption}. In other words, $\phi$ represents the number of TT4 in \ourname, where a new TT4 is generated every time \textit{secure decryption} is attempted for $K \leq 1$ as seen in Step A of \sect\ref{high_level}. Note that we have simulated three attempts (Session 3, 5 \& 7) to access $W_1$, which forces the Defender to adjust $R_C$. In particular, we observe that $R_C$ is severely affected by $\phi$ in comparison to the number of training sessions ($s$). This indicates that to increase $R_C$, the contract must behave normally for a large number of training sessions such that the ratio $\phi/s$ approximates 0.\\
\textbf{Effect of PMR in $R_\tau$.} For this experiment, we vary the PMR to $\{0.1,0.3,0.5\}$ to observe the behavior of the training reward $R_\tau$. Fig.~\ref{fig:rewards}b shows the benign $R_\tau$ and malicious $R_\tau$ under different PMR levels. 
Note that the process to determine $R_\tau$ for each client is based on the number of malicious models found during a training round. In other words, the Defender forces malicious clients to transfer their potential rewards to benign ones, i.e., benign clients get 140\% (Round 8), 160\% (Round 4) and 200\% (Round 3) for PMR of 0.1, 0.3 and 0.5, respectively. This process promotes benign training behavior since $R_\tau$ is reduced to 0 for every client detected as malicious during a training round.

\section{Security and Privacy Analysis}
The following sections provide a security and privacy analysis to further explore the resiliency against white-box inference and poisoning attacks under different adversary configurations. We also discuss the robustness of \ourname against clients randomly dropping from the learning process in App.~\ref{app:droupouts}.
\subsection{\ourname Privacy Analysis}
\ourname uses a decentralized crypto-system maintained by Gateway and Defender contracts. This allows \ourname to manage and analyze ciphers. \ourname adopts a semi-honest security model such that only one (out of 2) entity could be compromised at the time as discussed in assumption \textbf{A2}. 
Therefore, to undermine the privacy of \ourname according to \sect\ref{sect:adversary-privacy}, an adversary \adversaryPrivacy may formulate one of the following scenarios. \\
\textbf{\adversaryPrivacy Compromises Gateway Contract.} If \adversaryPrivacy maliciously controls the Gateway contract, \adversaryPrivacy would have access to every encrypted model. However, \adversaryPrivacy cannot decrypt any model directly as the private key is only held by the Defender contract. To access local updates, \adversaryPrivacy may try the following strategies.\\
\emph{Direct Decryption Request.} \adversaryPrivacy directly requests decryption of encrypted models by submitting the appropriate ciphers to Defender. This initial approach yields ineffective results as the decryption process follows \textit{secure decryption} (Step A in \sect\ref{design}),     where this process can identify the type of computation (i.e., summations or model operations) by analyzing data composition after decryption. To mitigate this attack, our approach identifies each cipher as a model operation and returns decrypted arrays with injected random noises $\delta$. Consequently, attempts to decrypt local updates result in random shifts to the distribution of each model. Therefore, this defense can effectively obfuscate local updates when an adversary attempts to decrypt them directly.\\
\emph{Reverse Engineer from Computations.} A sophisticated \adversaryPrivacy may consider to reverse engineer encrypted local updates from the result of specific computations, i.e., $G^*_t + \delta^*_i - W^*_i$, ${W^*_i}^2$. However, this approach is also found ineffective since any type of model operation is constrained by \textit{secure decryption}, resulting in data arrays being indirectly affected by $\delta$.\\
\emph{\ourname Aware Decryption Request.} \adversaryPrivacy attempts decryption of the first ($K=1$) encrypted model committed to \ourname during a training round to bypass the security measures imposed by \textit{secure decryption}. Put differently, \adversaryPrivacy aims to extract the first local model before other $\delta$ values skew the results. However, this approach is also ineffective because the Defender contract is aware of the number of models currently present in \ourname. As a consequence, the Defender contract leverages that information to adjust the contract reward $R_C$ to penalize the attempt and address curious behavior as illustrated in \sect\ref{sec:reward_system}.    

\noindent\textbf{\adversaryPrivacy Compromises Defender Contract.} For the following scenario, \adversaryPrivacy attempts to visualize local weights during \textit{secure decryption} as Defender holds the private key. To achieve this goal, \adversaryPrivacy requires the assistance of Gateway contract as the latter is the one that holds every encrypted model. Put differently, \adversaryPrivacy needs the Gateway to send encrypted models rather than encrypted computations, which breaks assumption \textbf{A2}. Therefore, \ourname is resistant to \adversaryPrivacy given assumption \textbf{A2} holds. \\ 
\textbf{\adversaryPrivacy Under Limited Collaboration.} In this scenario, \adversaryPrivacy is aware of \ourname's limitation and aims to retrieve the first local model ($K=1$) as described in \sect~\ref{sec:eval-miaMitigation}. However, our evaluation showed that \ourname is also resilient to this scenario, since \adversaryPrivacy's probability of \mbox{success reaches $\sim$0\% as defined by $1/m!$.}

\subsection{\ourname Security Analysis}
\ourname  efficiently filters \sota backdoor injections with the defense deployed in Defender contract. To bypass \ourname, an adversary \adversaryPoisoning should inject a poisoned model such that \ourname cannot distinguish benign models from poisoned ones. The following elaborate the methodologies that could be used by \adversaryPoisoning to achieve this goal.\\
\textbf{\adversaryPoisoning manipulates $\alpha$.}  \adversaryPoisoning continuously monitors and adjusts client's loss function to reduce the distance (i.e., cosine or L2 norm) between the client model and the current global model, a larger value of $\alpha$ means more aggressive attacks. \sect{sec:eval-backdoorMitigation} demonstrates that \ourname can eliminate poisoning attempts efficiently under different values of $\alpha$ for WP and IC applications, respectively. \\
\textbf{\adversaryPoisoning manipulates PMR.} \adversaryPoisoning would minimize its visibility by increasing the level of control over (or the number of)  malicious clients.  However, this approach cannot defeat \ourname as we empirically proved that \ourname maintains high performance w.r.t. changes in PMR for WP (App.~\ref{app:results_wp}) and IC learning tasks (App.~\ref{app:results_ic}).\\
\textbf{\adversaryPoisoning manipulates PDR.} \adversaryPoisoning can also limit the number of poisoned samples by decreasing the PDR value during training to generate less suspicious models. As a result, backdoors (BA) will be reduced. Additionally, regardless of PDR, \ourname continuously filter poisoned models efficiently as shown in App.~\ref{app:results_ic}. 

\vspace{-0.25em}
\section{Related Work}
\textbf{Privacy-Preserving Defenses.} Multiple approaches have been proposed to protect the privacy of the clients' training data. Passerat \etal use a blockchain for privately aggregating the individual models~\cite{Passerat}. Ryffel~\etal propose a framework that eases the use of Secure-Multi-Party Computation (SMPC) for secure aggregation~\cite{ryffel2018generic}, while Fereidooni~\etal rely only on SMPC~\cite{fereidooni2021safelearn}. Sav~\etal use Multiparty Homomorphic Encryption (HE) for collaboratively training a model~\cite{sav2020poseidon}. Bonawitz \etal propose a multi-party-compu- tation scheme based on Shamir's secret sharing~\cite{bonawitz2017practical}.  However, as this approach prevents the server from analyzing the local models, it also prevents analyzing them to identify poisoned models. \ourname uses Blockchain Two Contract Computation (BT2C) to engage in decentralized privacy-preserving computations based on smart contracts and HE (see Step A in \sect\ref{design}). \ourname raises accountability by including a reward system that promotes benign contract behavior (see evaluation in \sect\ref{sec:eval-miaMitigation} and \sect\ref{sec:reward_system}).\\
\textbf{Poisoning Defenses.} Existing defenses against data and model poisoning attacks aim at distinguishing malicious and benign model updates~\cite{kim2022backdoor,wang2022rflbat} utilizing filter-based approaches (i.e., clustering techniques). However, all of these defenses make specific assumptions about the distribution of benign and malicious data or the characteristics of injected models, causing them to fail if any of these assumptions do not hold. Moreover, such defenses cannot detect stealthy attacks, e.g., where the adversary constrains their poisoned updates within benign update distribution such as constrain-and-Scale attack~\cite{bagdasaryan2020backdoor}. Yin \etal~\cite{yin2018byzantine} and Guerraoui \etal~\cite{guerraoui2018} propose to change aggregation rules to mitigate the effect of malicious model updates. They utilize median parameters from all local models as the global model parameters. 
Other approaches validate the local models or the aggregated model~\cite{stripelis2022performance,zhao2020shielding,xi2021batfl}. However, they cannot detect stealthy backdoors that have only minimal impact on the Main Task Accuracy (MA). Weak Differential Privacy (DP) techniques~\cite{naseri2020local,xie2021crfl,miao2022against} have also been used to mitigate the effects of poisoning attacks. DP-based defense dilutes the impact of poisoned models by clipping model weights and adding noise to individual clients' model updates. DeepSight provided an efficient filtering that works even in \nonIId data scenarios~\cite{rieger2022deepsight} but does not credit the individual contributions and requires a central server than can inspect the model updates. Kalapaaking \etal use smart-contract to verify the client-side training process~\cite{kalapaaking2023smart} but cannot prevent attackers from manipulating the training data~\cite{nguyen2020poisoning}. The use of blockchain implementations~\cite{zhou2020pirate, li2020blockchain, desai2021blockfla,hua2020blockchain} have managed to provide defenses against poisoning attacks, however, these solutions only consider untargeted attacks and/or rely on specific assumptions about the distribution of training data. In contrast, \ourname effectively removes poisoned models by instantiating a filtering approach based on Gaussian Kernel Density Estimation (G-KDE) function (see Step 4 in \sect\ref{design}). The poisoning resiliency of our solution is empirically evaluated on \sect\ref{sec:eval-backdoorMitigation}, which demonstrates that \ourname is an effective solution to mitigate poisoning attacks. In addition, our solution allows us to treat the underlying problem of poisoning attacks, i.e., client training misbehavior, by imposing monetary deterrents for detected poisoning attempts. This is illustrated in \sect\ref{sec:reward_system}. \\
\textbf{Hybrid Defenses.} A number of existing works recently focused on poisoning attacks while preserving the privacy of the individual model updates. Two works implemented Krum~\cite{blanchard2017machine} using SMPC~\cite{khazbak2020mlguard, fereidooni2021safelearn}. However, Krum focuses on untargeted poisoning attacks and fails to effectively mitigate backdoor attacks (see \sect\ref{sec:eval-backdoorMitigation}), and SMPC is vulnerable to attacks that limit availability (i.e., DoS attacks). Similarly, FLAME~\cite{nguyen2022flame} leverages DP with model filtering to provide an efficient defense against backdoors. However, FLAME also relies on SMPC to provide privacy-preserving computations. Several approaches utilize Trusted Execution Environments (TEE) to realize a privacy-preserving poisoning detection~\cite{hashemi2021byzantine,mo2021ppfl,mondal2021flatee}. However, requiring TEEs restricts their application to a few scenarios as mobile devices often do not have a TEE, while \ourname does not make any assumption about the hardware. Biscotti~\cite{shayan2020biscotti}, BEAS~\cite{mondal2022beas} and the work of Lu~\etal\cite{lu2019blockchain} are blockchain-based frameworks that target secure and private FL. The first two approaches use Multi-Krum \cite{blanchard2017machine} (a Krum variant) to reduce the impact of backdoors in the system, where Multi-Krum also focuses on untargeted poisoning. In particular, Biscotti uses DP and Shamir secrets to preserve privacy of local updates during aggregation. BEAS and the work of Lu\etal, on the other hand, leverages DP to obfuscate model weights and provide resiliency against inference attacks. In comparison, \ourname provides~ (1) strong privacy guarantees to client models by obfuscating them using our BT2C protocol (see \sect\ref{sec:eval-miaMitigation}), (2) an effective defense using G-KDE functions to filter different poisoning attacks (see \sect\ref{sec:eval-backdoorMitigation}), and (3) compensation algorithms to automatically adjust the reward and deter malicious behavior (see \sect\ref{sec:reward_system}). Liu \etal proposed a smart-contract-based FL framework that utilizes a server-side validation dataset to detect untargeted poisoning attacks~\cite{liu2020secure}. However, assuming a server-side dataset is not practical~\cite{rieger2022deepsight}, while \ourname detects poisoning attacks without making such an assumption and also includes a reward mechanism to penalize malicious clients.
\vspace{-0.03cm}
\section{Conclusion and Future Work}
In this paper, we illustrated the current research gaps facing FL systems in terms of security and privacy. To fill in those gaps, we propose \ourname, a 3-layer blockchain FL framework powered by smart contracts and HE. Our proposed HE infrastructure, BT2C, enables the analysis and operation of model weights in cipher text through the use of a semi-honest collaboration between smart contracts. BT2C is shown to provide resiliency against white-box inference attacks with a decreased adversarial probability of success of $\frac{1}{m!}$, where $m$ is the number of ciphers inside an encrypted model. Furthermore, our extensive evaluation shows that \ourname also offers high resiliency against numerous poisoning strategies (BA$\approx$0) with minimal impact on benign accuracy. Our solution to both white-box inference and poisoning attacks allows us to effectively embed incentives or penalizations to both aggregation contracts (e.g., Gateway) and training clients in an effort to minimize adversarial intent via behavior accountability. \\
\textbf{Limitations.} Although the use of blockchain technology and HE enable the security properties in \ourname, they also contribute to the growth in computation costs of its learning process. For instance, training round latency increases from 15.86s (MNIST model) to 76.6s (CIFAR10 model). This indicates that \ourname has an increase in latency of approximately five-times (4.82) for a model that is ten-times larger (10.17). Put differently, \ourname offers limited scalability given that its learning process slows down as both model complexity and the number of clients/models increases in the system. Furthermore, the reward mechanism embedded in \ourname is directly related to the performance of its defensive strategies. For example, an attacker capable of avoiding \ourname's poison defense (i.e., G-KDE Defense) may continue to receive credit even though its model negatively contributes to the learning process.  \\ 
\textbf{Future Work.} A formal in-depth analysis aimed into the scalability of \ourname, e.g., transaction fees, storage costs, computation costs and communication costs, is needed to generate additional insights into the limitations and efficiency of \ourname, specifically those imposed by the use of different blockchain platforms. 
\vspace{0.05cm}
\section*{Acknowledgment}\vspace{-0.05cm}
This work was supported in part by the U.S. Department of Energy/National Nuclear Security Administration (DOE/NNSA) \#DE-NA0003985, Intel through the Private AI Collaborate Research Institute (\url{https://www.private-ai.org/}), BMBF and HMWK within ATHENE, as well as from the OpenS3 Lab, the Hessian Ministry of Interior and Sport as part of the F-LION project, following the funding guidelines for cyber security research, the Horizon Europe framework program of the European Union under grant agreement No. 101093126 (ACES). We extend our appreciation to KOBIL GmbH for their support and collaboration throughout the course of this project. Any opinions, findings, conclusions, or recommendations expressed in this material are those of the authors and do not necessarily reflect the views of any of these funding agencies.

%\printbibliography
\bibliographystyle{ACM-Reference-Format}
\bibliography{main.bib}
\vspace{-0.15cm}
% % --- Appendix ---%
\appendix\vspace{-0.15cm}
\section*{Appendix}

\section{Background on Federated Learning}\label{app:fl_background}
Federated Learning (FL) \cite{mcmahan2017communication} is a round-based machine learning protocol with the purpose of providing a collaborative learning surface between $K$ clients and a central aggregation server \aggregationServer. For a training round $t \in [1,T]$ in \fedavg \cite{mcmahan2017communication}, each client downloads an initial deep learning model from \aggregationServer known as the global model $G_{t-1}$. Clients begin training using its local data $D_i$ and send local updates $W_i$, where $i \in [1,K]$ to be averaged by \aggregationServer resulting in a new global model $G_t$, such that $G_t = \sum_{i=1}^{K} \frac{n_i}{n}W_i$, where $n_i=\| D_i \|$, and $n=\sum_{i=1}^K n_i$. The process continuously repeats for $T$ rounds until the model performance reaches an appropriate/target level. Similar to previous work \cite{bagdasaryan2020backdoor,shen2016auror}, we weight all clients equally such that $G_t = \sum_{i=1}^{K} \frac{W_i}{K}$.

\section{Background on Blockchain} \label{app:block_background}
Blockchain is a decentralized storage system sustained over a vast network of computers (peers) with the common goal to serve as an immutable record \cite{nakamoto2019bitcoin}. The record is formed from a series of cryptographically linked \textit{blocks}, where each block contains multiple transactions (or data samples). The transaction order is defined by a systematized ordering protocol. The protocol is also known as the consensus process (i.e., proof of work (PoW) \cite{gervais2016security} or proof of stake (PoS) \cite{king2012ppcoin}) that uses a subset of peers to approve and validate new blocks of data. Informally, the blockchain participants broadcast data received by consensus nodes as transactions. These nodes gather different transactions to formulate a new block during the consensus process. After consensus is reached, the block is distributed to every other participant so that it can be validated and appended to the ledger. 

Smart contracts of blockchain systems allow authorized users to retrieve any data from and submit any data to the underlying blockchain system. They provide a verifiable and transparent environment dedicated to build trust among untrusted parties without an intermediary. The use of smart contracts in the FL context enables a decentralized storage to merge and analyze client models in an automated fashion. Currently, blockchains can be categorized as either a public or private.\footnote{Public (or permissionless) blockchains allow anyone to join and verify the system, whereas private (or permissioned) blockchains do not.} For our solution we consider a private blockchain to have additional control over the visibility of the ledger as only specific entities are selected to join the network. 
\textbf{Hyperledger Fabric} (HLF) offers high degree of flexibility and scalability~\cite{androulaki2018hyperledger} while utilizing a modular approach that is easy-to-use. We use HLF as the blockchain platform in this study because of the following reasons.

\begin{itemize}
    \item \textit{Scalability.} Compared with public blockchains that require every consensus node to validate newly appended blocks, Fabric controls the number of consensus nodes to improve its throughput and significantly reduce the communication latency.
    \item \textit{Leader-based consensus process.} HLF uses Raft \cite{ongaro2015raft}, a fault-tolerant algorithm that requires nodes to elect a leader to process and distribute data. This leader-based approach reduces data processing times and makes the blockchain systems time/computational efficient.
\end{itemize}

\section{Background on Homomorphic Encryption}\label{app:he_background}
Homomorphic Encryption (HE) \cite{yi2014homomorphic, acar2018survey} is a special public key encryption. It allows the application of mathematical functions on encrypted data so that data can remain confidential while being processed. HE has two main branches, namely fully homomorphic encryption (FHE) and partially homomorphic encryption (PHE). FHE allows users to compute \textit{any} mathematical function while PHE only allows \textit{some} functions to be applied~\cite{chen2019improved, fan2012somewhat}. Prominent examples of PHE are Cheon-Kim-Kim-Song (CKKS)~\cite{cheon2017homomorphic} and Brakerski/Fan-Vercauteren (BFV) \cite{fan2012somewhat} schemes. \\
\textbf{CKKS Encryption Scheme.} Computing only addition and multiplication operations is a prominent example of practical PHE \cite{chen2019improved}. Unlike other PHE (i.e., BFV \cite{fan2012somewhat}), CKKS allows floating point operations, which makes it applicable for encrypting data and models for FL systems. In typical CKKS implementations, the encryption process begins when a client first generates a key pair, a secret-key $S_k$ and a public-key $P_k$. The client uses $P_k$ to encrypt its data $D_i$ before sharing it to an untrusted server $S_u$ to perform private computations on encrypted data $D^*_i$. Computed data is then returned to the client for decryption with $S_k$, unveiling the real result after computation. Previous research~\cite{naehrig2011can,bos2014private} demonstrated the practicality of HE in applications where privacy is preferred over efficiency.

\section{Intuition for G-KDE Clustering}\label{app:intuition}
\begin{figure}[!t]
\centering
\includegraphics[width=0.85\columnwidth]{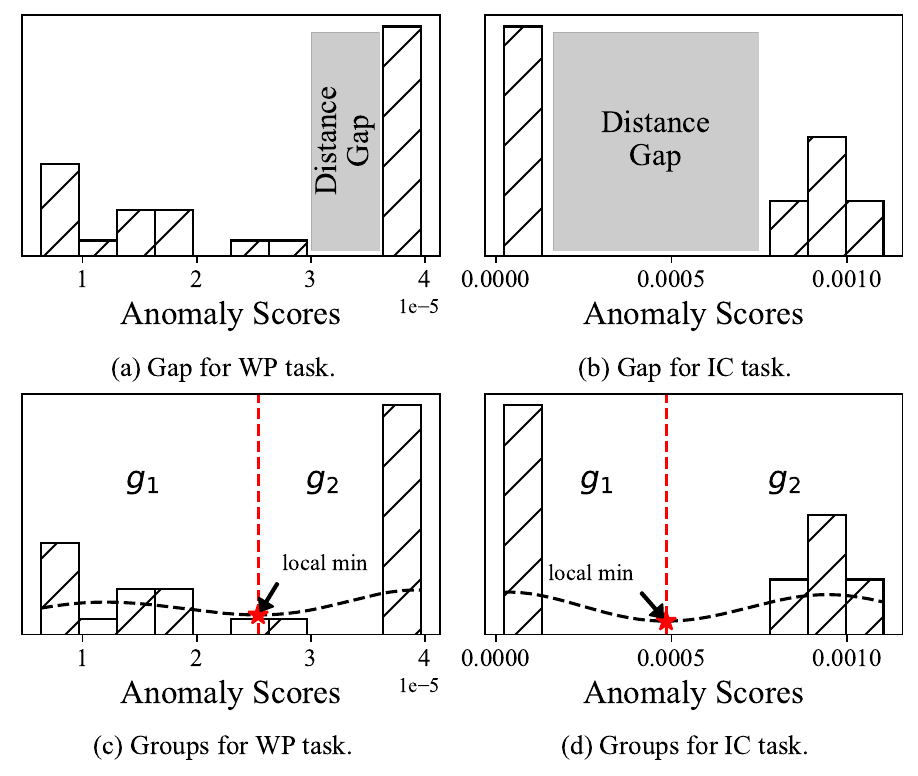}
\caption{Distribution analysis of anomaly scores for WP and IC datasets.  (a) A gap between malicious and benign models for WP. (b) A gap between malicious and benign models for IC. (c) Groups found during WP analysis. (d) Groups found during IC analysis.}
\label{fig:dist_cluster}
\end{figure}

Poisoning attempts are known to generate a sufficiently large gap between the distance scores of benign models and malicious updates such that our G-KDE clustering is able to filter malicious models regardless of the obfuscation technique used by an adversary (e.g., manipulation of loss function \cite{bagdasaryan2020backdoor}). This hypothesis is validated and shown in Fig.~\ref{fig:dist_cluster}, where four subfigures illustrate the distance score distribution found for both word prediction tasks (WP) on the left, and image classification tasks (IC) on the right. Fig.~\ref{fig:dist_cluster}a uses a histogram to prove the existence of a gap within distance scores of WP tasks. Similarly, Fig.~\ref{fig:dist_cluster}b displays a similar gap between the malicious and benign scores obtained from an IC task. These gaps indicate that it is possible to discriminate models based on where they reside within the data distribution regardless of model complexity, making this approach generic enough to be applicable as a poisoning defense. Fig.~\ref{fig:dist_cluster}c and d illustrate how the use of G-KDE functions finds multiple distributions inside the distance scores, thus, generating two groups, e.g., $g_1$ (benign) and $g_2$ (malicious).

\section{Evaluation Metrics}
\label{app:metrics}
We consider the following metrics to define the effectiveness of poisoning attacks:
\begin{itemize}
    \item \textit{Main Task Accuracy (MA).} It indicates the accuracy level of a model when performing its main (or benign) task. The goal of an adversary is to minimize its impact such that MA is not degraded during malicious training.
    \item \textit{Backdoor Accuracy (BA).} It is used to measure the accuracy of the backdoor injected into a model. The goal of the attacker is to maximize BA during a training round. 
    \item \textit{True Positive Rate (TPR).} It indicates the accuracy of the defenses for detecting poisoned models such that True Positives (TP) are models correctly identified as malicious, and False Negatives (FN) are those mislabeled as benign: $TPR = \nicefrac{TP}{TP + FN}$ . A high TPR suggests poisoned models are being removed:

    \item \textit{True Negative Rate (TNR).} It determines the accuracy of the defense for detecting benign models such that True Negatives (TN) are models correctly labeled as benign, and False Positives (FP) are those incorrectly classified as malicious: $TNR = \nicefrac{TN}{TN+ FP}$. TNR increases as the defense removes less benign models. 

\end{itemize}

\section{Runtime-Performance}\label{app:performance}
\begin{figure}[b]
    \centering    
    \includegraphics[width=0.6\columnwidth]{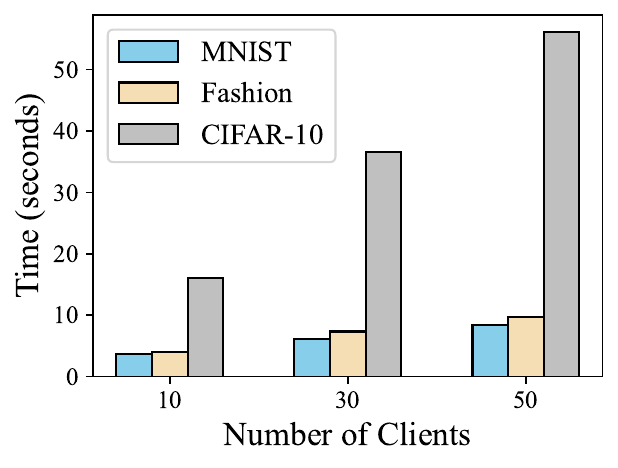}
    \caption{Effect of number of training clients on Model Aggregation process.}
    
    \label{fig:agg_process}
\end{figure}

To assess the latency aspects of \ourname during any given round, we evaluate the following processes in terms of latency: model encryption, model upload, model filtering and model aggregation. Model encryption determines the time required to fully produce an encrypted model. Model upload defines the average time an encrypted model is committed and analyzed by \ourname such that it produces the distance score that corresponds to the uploaded encrypted model. Model filtering evaluates how long the poison defense takes to analyze anomaly scores to remove models from aggregation. Model aggregation determines the time it takes to produce a new global model. We perform experiments using MNIST, Fashion, and CIFAR-10 datasets. \\ 
\textbf{Effect of Model Complexity.} For the following experiment we set the number of clients to be 50 to assess the impact of model complexity on system latency. Table~\ref{tab:latency} compiles the preliminary evaluation of \ourname. 
For model encryption, we found there is negligible impact on latency with values of 0.12, 0.14, 0.56 and 1.03 seconds for each of the models evaluated. Note that model encryption behaves as $O(n^2)$ as the number of trainable parameters increases, i.e., $\sim$ 20.4M params are encrypted (10K ciphers) in $\sim$110s. Moreover, model upload shows an average latency of 5.23, 5.85, and 17.57 seconds for MNIST, Fashion, and CIFAR-10 tasks. The increased latency is mainly attributed to the distance score computation, where we use three BT2C rounds to compute the cosine distance. The values displayed for model filtering are close to 2s in each evaluated task. The reason is because this is the only process where HE is not used, as the Defender contract only analyzes the anomaly scores previously computed during model upload. In contrast, model aggregation is determined to be the most computationally expensive process in \ourname, showing values of 8.44, 9.73, and 56.09 seconds, respectively. As a consequence, \ourname successfully completes a single training round in 15.86s (MNIST), 17.7s (Fashion), and 76.6s (CIFAR-10). Therefore, the use of \ourname is best suited to protect the privacy of constrained environments, where local clients do not have the computational resources (e.g., GPU) to train a robust model with millions of parameters such as ResNet \cite{he2016deep} and VGG \cite{simonyan2014very}. However, they can operate lite models, i.e., lite-CNN and/or ConvMixer, to achieve good model performance while delegating every intensive computational task to blockchain via smart contracts. \\
\textbf{Effect of Number of Clients.} In this experiment, we narrow our focus to the model aggregation process since it was determined to be the most computationally expensive process in \ourname. To further inspect this process, we vary the number of clients to $\{10, 30, 50\}$. Fig.~\ref{fig:agg_process} shows the impact of model aggregation for the distinct learning tasks using a different number of clients. Here, we can visualize that not only does model aggregation scales according to model complexity, but it also shows a linear dependency w.r.t. the number of clients. Put differently, model aggregation is the major contributor of latency in \ourname as other components (i.e., model encryption, model upload) are only defined by the computational requirements of HE.  

\begin{table}[!t]
\caption{Training round efficiency of \ourname for the following processing tasks in seconds: Model Encryption, Model Upload, Model Filtering, and Model Aggregation.}
\label{tab:latency}
\centering
     \begin{tabular}{r|c|c|c}
    \hline
    \textbf{Dataset} & \textbf{MNIST} & \textbf{Fashion} & \textbf{CIFAR-10} \\
    \hline
    \textbf{Model Encryption} & 0.12 & 0.14 & 1.03\\
    \textbf{Model Upload} & 5.23 & 5.85 & 17.57\\
    \textbf{Model Filtering} & 2.07 & 1.98 & 1.91\\
    \textbf{Model Aggregation} & 8.44 & 9.73 & 56.09\\
    \hline
    \textbf{Total} & 15.86 & 17.7 & 76.6\\
    
    \end{tabular}
    
\end{table}

\section{Robustness Against Client Dropouts}
\label{app:droupouts}
 The goal of the training client is to train and submit its local update to the Gateway contract. Uploading the model can be seen as an automatic operation, either the model is completely received and further processed or it is ignored. Once a local model is successfully uploaded, a TT2 is formulated to generate a record in the ledger. At this stage, the model is now considered to be part of \ourname. Otherwise, the encrypted model is ignored during the training round. After the record for a model is part of the ledger, a client dropout does not affect \ourname anymore. \ourname takes advantage of the visibility of blockchain to consistently account for the number of encrypted models currently present. This allows our approach to perform the aggregation using only the available models. Thus, making it robust against dynamic dropouts.

 \section{Backdoor Evaluation for WP Task}\label{app:results_wp}
\begin{figure}[t]
    \centering    
    \includegraphics[width=0.7\columnwidth]{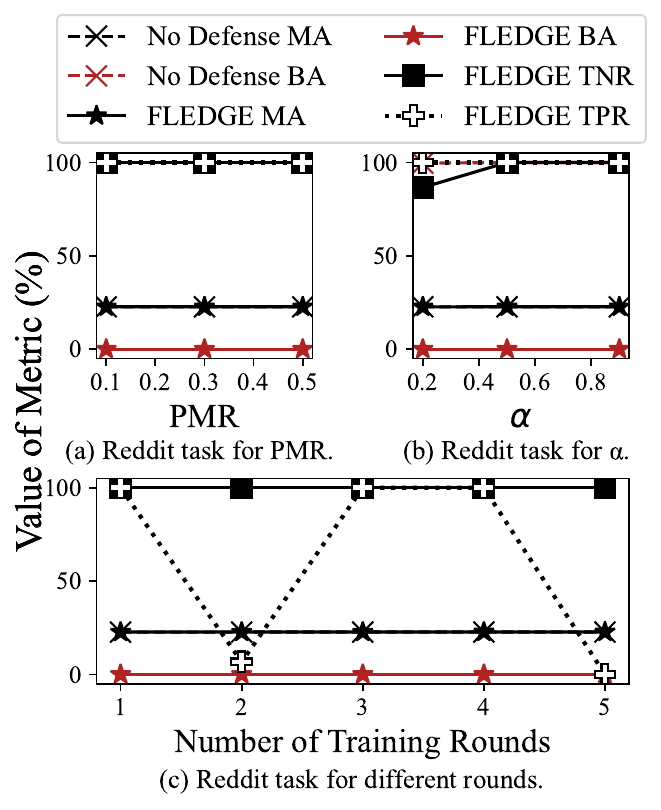}
    \caption{Effect of backdoors on evaluation metrics for WP task. Effect of (a) PMRs, (b) $\alpha$, (c) multiple injections.}
    \label{fig:wp}
\end{figure}

The following section evaluates the effect of PMR and $\alpha$ in \ourname for the WP task. We analyze the behavior of \ourname during 10 training rounds. The attacker attempts to inject backdoored models in different training rounds to compromise the global model. We analyze \ourname in terms of MA, BA, TPR and TNR, and draw a comparison with a typical \textit{No Defense} scenario using a FL environment with 30 training clients. The results of the direct comparison are shown in Fig.~\ref{fig:wp}, and are elaborated as follows. \\
\textbf{Effect of PMR Rate.} In this experiment, we modify the control ratio of the attacker in the system by setting PMR values of $\{0.1, 0.3, 0.5\}$ (or 6, 9 and 15 malicious clients). In Fig.~\ref{fig:wp}a, we observe the ability of the adversary to successfully inject backdoors (BA=100\% for \textit{No Defense}) into the model regardless on the number of malicious clients it controls. In contrast, \ourname is demonstrated to effectively remove the threat of backdoors (BA=0\%) while preserving the benign accuracy of MA=22.7\%. \ourname achieves a TNR and TPR values of 100\% in the process, which denotes that \ourname is able to defend against multiple malicious clients. \\
\textbf{Effect of $\alpha$ Rate.} Further, we modified the level of intensity for backdoor injection such that $\alpha$=$\{0.2,0.5,0.9\}$. Fig.~\ref{fig:wp}b shows that BA and MA (\textit{No Defense}) are not affected by $\alpha$ with an average value of 100\% and 22.6\%, respectively. \ourname effectively mitigates every backdoor attempt (BA=0\%, TPR=100\%). TNR is 86.66\% when $\alpha$=0.2, and this value increases to 100\% when $\alpha$=$\{0.5,0.9\}$. \ourname filters 2 out of 15 (13.34\%) benign models when defending against $\alpha$=0.2, and it continues to provide an effective defense against different \mbox{levels of aggression when $\alpha$ value increases.} 
 ~\\
\textbf{Effect of Backdoor Injections for Multiple Rounds.} For this experiment, we set PMR=0.5 and the number of rounds to be 5, where rounds 1,3 and 4 are poisoned, and rounds 2 and 5 are benign. This is because we are interested in testing the efficacy of \ourname w.r.t. multiple attack rounds. Fig.~\ref{fig:wp}c illustrates the operation of \ourname under multiple attack rounds. Note that for all 5 rounds, \ourname continues to maintain BA=0\%, whereas the \textit{No Defense} setup has BA=100\%. Further, TNR values are always 100\%, however, TPR values differ depending if the round is benign (TPR$\approx$0\%) or malicious (TPR=100\%). This behavior indicates that \ourname adjusts its filtering mechanics to only remove poisoned models and avoid benign models.

\section{Backdoor Evaluation for IC Task}\label{app:results_ic}
\begin{figure}[t]
    \centering
    \includegraphics[width=0.9\columnwidth]{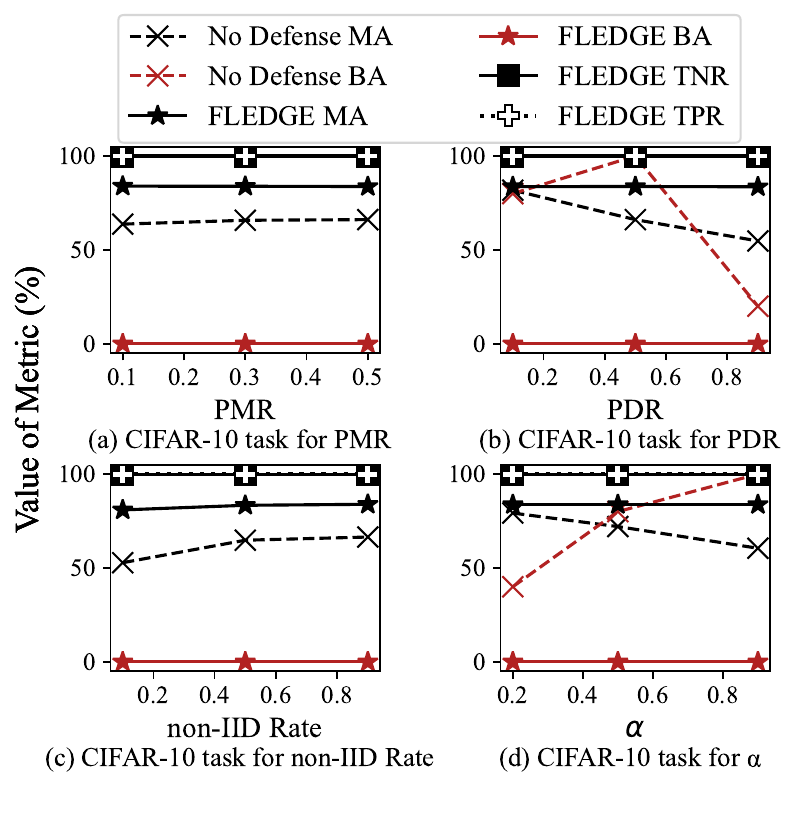}
    \caption{Effect of backdoors on evaluation metrics for IC task. Effect of (a) PMRs, (b) PDRs, (c) non-IID rates, (d) $\alpha$.}
    \label{fig:ic}
\end{figure}

In this section, we evaluate the effect of PMR, PDR, non-IID rate and $\alpha$ on IC tasks during a single training round. We illustrate the difference in behavior for \ourname and \textit{No Defense} system during multiple attack settings. We use MA, BA, TPR, and TNR as metrics for comparison. For the purposes of demonstration, we focus on CIFAR-10 as it is the most complex IC task. 
We select 50 as the total number of clients. The results of the \mbox{experiments are illustrated in Fig.~\ref{fig:ic}.}\\
\textbf{Effect of PMR Rate}. The following experiment changes the influence of the attacker over the system by setting PMR values of $\{0.1, 0.3, 0.5\}$. Fig.~\ref{fig:ic}a shows the evaluation for PMR. At the \textit{No Defense} setup, we clearly observe the attacker negatively affects MA from 83.9\% to 66.1\% to reach an appropriate BA level (BA=100\%). Nevertheless, \ourname efficiently mitigated this, since \ourname continues to filter malicious updates (BA$\approx$0\%), reaching TNR and TPR of 100\%. Consequently, this elevates MA to a benign level, i.e., MA=83.94\%. Therefore, we further demonstrate that \ourname is resistant to backdoor injections where the adversary is able to change the number of clients it controls. \\
\textbf{Effect of PDR Rate.} In this experiment, we evaluate the influence of PDR in \ourname by setting PDR=$\{0.1, 0.5, 1\}$ as illustrated in Fig.~\ref{fig:ic}b. Similar to previous experiments, Fig.~\ref{fig:ic}b shows the evaluation metrics for different PDR values. Here, we observe the following BA behaviors. First, it shows a mild drop in performance for PDR=0.1, i.e., BA=80\% and MA=81.96\%. Second, it shows a significant degradation in performance when PDR=1, i.e., BA=20\% and MA=54.8\%. This is, however, filtered by \ourname which effectively reduces BA to 0\% with TNR=100\% and TPR=100\%. Therefore, we consider \ourname to be resistant to changes in PDR.\\
\textbf{Effect of non-IID Data.} For the following experiment, we aim to analyze \ourname under different data concentrations such that non-IID is set to $\{0, 0.5, 1\}$. Fig.~\ref{fig:ic}c show that backdoors continue to be effective for every non-IID value, with BA of 100\%. These results also show that MA increases from 52.8\% to 66.5\% for non-IID of 0 and 1, respectively. However, \ourname minimizes the impact of backdoors (BA$\approx$0\%) for every non-IID setting with TNR and TPR of 100\%. This indicates that \ourname is highly resilient to changes caused by different non-IID rates. \\ 

\textbf{Effect of $\alpha$ Rate.} For the next experiment, we set $\alpha$ to be $\{0.2,0.5,0.9\}$ in order to test the performance of \ourname under different intensity levels. 
Similar to previous experiments, Fig.~\ref{fig:ic}d shows the evaluation metrics. In here, we observe how BA is directly proportional to $\alpha$, i.e., BA of 40\% ($\alpha$=0.2) increases to 100\% ($\alpha$=0.9). Consequently, this also has an impact on MA, yielding reduced values of 60.5\% when $\alpha$=0.9. \ourname removed poisoned models (TPR=100\%) while preserving benign ones (TNR=100\%) such that BA=0\% and MA is returned to its benign level, i.e., 83.8\%. Hence, we further determine that \ourname is robust against changes in~$\alpha$.
\end{document}